\documentclass[journal]{IEEEtran}
\usepackage{graphicx}
\usepackage[utf8]{inputenc} 
\usepackage[T1]{fontenc} 
\usepackage{amsmath}
\usepackage{amsmath}
\usepackage{caption}
\usepackage{amssymb}
\usepackage{lettrine}
\usepackage{tikz}
\usepackage{verbatim}
\usepackage{cite}
\usepackage{array}
\usepackage{makecell}
\usepackage{chronosys}
\usepackage{ mathrsfs }
\usetikzlibrary{arrows,shapes,positioning,shadows,trees}
\usepackage[autostyle=true]{csquotes} 
\usepackage{bchart}

\begin{document}
\title {A Taxonomy of Data Attacks in Power Systems}
% \author{Sagnik~Basumallik,~\IEEEmembership{Student Member,~IEEE}
\author{Sagnik~Basumallik,~\IEEEmembership{Student Member, IEEE} \vspace{-1em}
	\thanks{This work is supported by the National Science Foundation (NSF) Grant No.1600058.}}
	
\maketitle

\begin{abstract}
In a macro-economic system, all major sectors: agriculture, extraction of natural resources, manufacturing, construction, transport, communication and health services, are dependent on a reliable supply of electricity. Targeted attacks on power networks can lead to disruption in operations, causing significant economic and social losses. When cyber networks in power system are compromised, time-critical data can be dropped and modified, which can impede real time operations and decision making. This paper tracks the progress of research in power system cyber security over the last decade and presents a taxonomy of data attacks.
Nineteen different attack models against major operation and control blocks are classified into four areas: steady state control, transient and auxiliary control, substation control and load control. For each class, a comprehensive review of mathematical attack models is presented. The goal is to provide a theoretically balanced approach to cyber attacks and their impacts on the reliable functioning of the electric grid.
\end{abstract}
%  This paper aims to highlight the details of different data injection attacks that can be launched against  

\begin{IEEEkeywords}
	cybersecurity, data attacks, smart grids
\end{IEEEkeywords}

\IEEEpeerreviewmaketitle

\section{Introduction}

The last few years have seen a surge in the number of cyber attacks emerging against industrial control systems. Examples include (a) `Industroyer', a malware that targeted switches, breakers and substation communication protocols, and was chiefly responsible for Ukraine's power grid failure \cite{CherepanovAnton,Liang2017TheAttacks}, (b) `Dragonfly', which allowed attackers to gain unauthorized control of critical systems \cite{2014Dragonfly:Response}, (c) `WannaCry', that affected operations of hospitals, banks and universities \cite{Chen2017AutomatedRansomware,Hsiao2018TheRansomware}, (d) `Stuxnet', which gained control of nuclear facilities \cite{Langner2011Stuxnet:Weapon,Karnouskos2011StuxnetSecurity,Chen2011LessonsStuxnet}, (e) `BlackEnergy', designed to launch denial-of-service attacks against SCADA applications in industrial control systems \cite{ThreatSTOP2016BlackEnergyReport} and (f) Trisis \cite{dragosTrisis}, a malware that attacked equipment used in energy, oil and gas control system. These threats result in substantial economic and social losses, and pose a significant challenge to the operation of time-critical systems like electric grids. This ushers in a rising need to understand attack models and their impacts if utilities are going to adopt preventive cyber-security schemes to protect critical infrastructures against data attacks.

Data attacks are defined as \textit{injection, alteration,  blocking, deletion, modification of data or a combination of any of the above, in devices or in communication network channels}, that impede the reliable operation of power systems. Such attacks can be either packet data injection attacks, denial of service attacks or time synchronization attacks launched through malwares, viruses, ransom-wares, fraudulent emails and social engineering \cite{ComputerMcAfee}. 

In this paper, a taxonomy is developed that classifies nineteen data attack models against four operation and control blocks of power system: (a) steady state control, (b) transient and auxiliary control, (c) substation control and (d) energy/load control, as shown in Fig \ref{fig:taxonomy}. Operation of control blocks and monitoring units susceptible to data attacks are first described. This is followed by discussion of mathematical attack models. Finally, the impacts of cyber threats on the physical grids are highlighted. In short, this paper brings together various research on data attack designs against power system components under one umbrella.

% Other particular surveys on data attacks include those against state estimation \cite{Deng2017}, electricity market \cite{Liang2017} and smart grid data communication \cite{Yan2013AChallenges}. 

\tikzset{
  basic/.style  = {draw, text width=3.5cm, drop shadow, font=\sffamily, rectangle},
  root/.style   = {basic, rounded corners=2pt, thin, align=center,
                   fill=green!30},
  level 2/.style = {basic, rounded corners=6pt, thin,align=center, fill=green!60,
                   text width=8em},
  level 3/.style = {basic, thin, align=left, fill=pink!60, text width=7em}
}
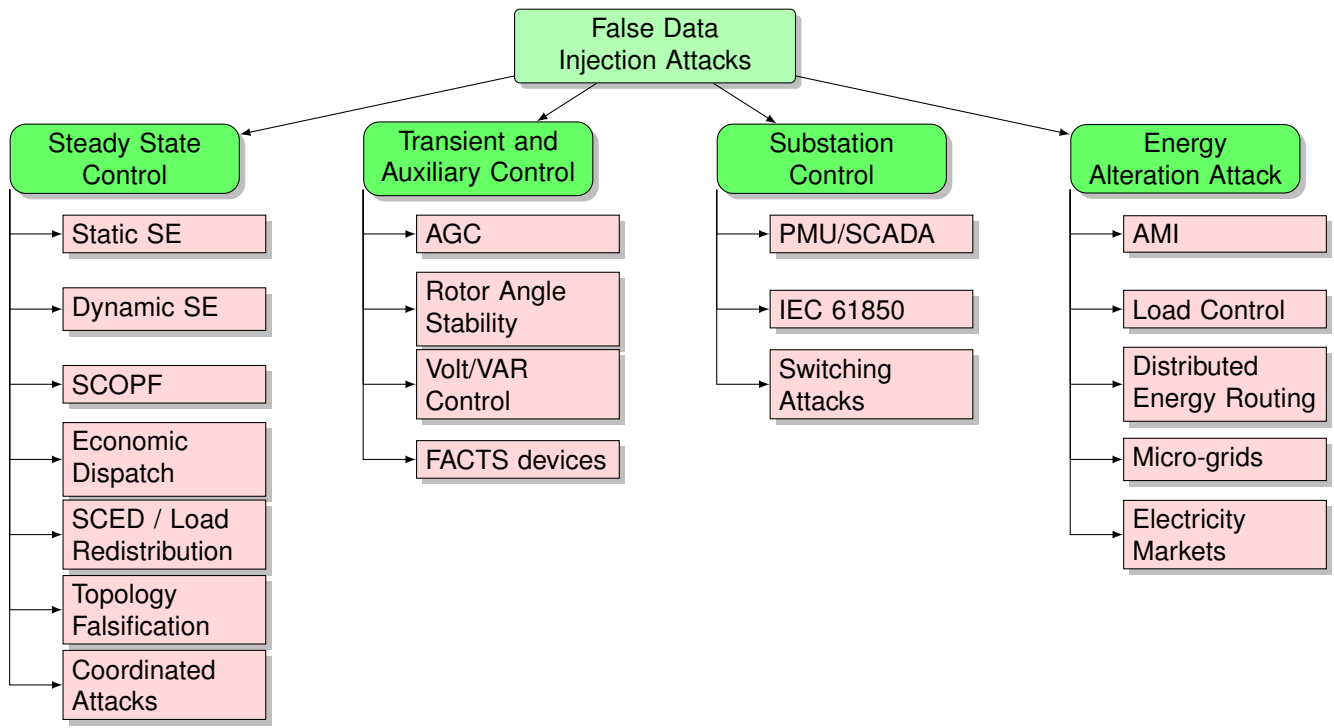
\begin{figure*}[htpb]
\centering
\begin{tikzpicture}[
  level 1/.style={sibling distance=47mm},
  edge from parent/.style={->,draw},
  >=latex]
Economic Dispatch
\node[root] {False Data Injection Attacks}
  child {node[level 2] (c1) {Steady State Control}}
  child {node[level 2] (c2) {Transient and Auxiliary Control}}
  child {node[level 2] (c3) {Substation Control}}
  child {node[level 2] (c4) {Energy Alteration Attack}};

\begin{scope}[every node/.style={level 3}]
\node [below of = c1, xshift=15pt] (c11) {Static SE};
\node [below of = c11] (c12) {Dynamic SE};
\node [below of = c12] (c13) {SCOPF};
\node [below of = c13] (c14) {Economic Dispatch};
\node [below of = c14] (c15) {SCED / Load Redistribution};
\node [below of = c15] (c16) {Topology Falsification};
\node [below of = c16] (c17) {Coordinated Attacks};

\node [below of = c2, xshift=15pt] (c21) {AGC};
\node [below of = c21] (c22) {Rotor Angle Stability};
\node [below of = c22] (c23) {Volt/VAR Control};
\node [below of = c23] (c24) {FACTS devices};

\node [below of = c3, xshift=15pt] (c31) {PMU/SCADA};
\node [below of = c31] (c32) {IEC 61850};
\node [below of = c32] (c33) {Switching Attacks};

\node [below of = c4, xshift=15pt] (c41) {AMI};
\node [below of = c41] (c42) {Load Control};
\node [below of = c42] (c43) {Distributed Energy Routing};
\node [below of = c43] (c44) {Micro-grids};
\node [below of = c44] (c45) {Electricity Markets};
\end{scope}

\foreach \value in {1,2,3,4,5,6,7}
  \draw[->] (c1.195) |- (c1\value.west);

\foreach \value in {1,...,4}
  \draw[->] (c2.195) |- (c2\value.west);

\foreach \value in {1,2,3}
  \draw[->] (c3.195) |- (c3\value.west);
  
\foreach \value in {1,...,5}
\draw[->] (c4.195) |- (c4\value.west);
\end{tikzpicture}
\caption{A taxonomy of data falsification attacks against various power system control and operation blocks}
\label{fig:taxonomy}
\vspace{-1em}
\end{figure*}

Some of the major steady state algorithms in power systems affected by data breaches include static state estimation (SSE), dynamic state estimation (DSE), optimal power flow (OPF) and security constrained economic dispatch (SCED). Such data attacks involve load measurement redistribution, topology falsification or a coordination of physical attacks augmented with false data. Apart from steady state operations, transient and auxiliary control blocks such as rotor angle stability, automatic generation control (AGC), automatic voltage control (AVR), volt/VAR controls and FACTS devices are also vulnerable to data attacks. Additionally, substation control and communication architecture like IEC 61850, PMU/SCADA data communication channels can also become potential  attack targets. Other targets include advanced metering infrastructure (AMI), residential load controls, distributed energy routing algorithms, micro-girds and electricity markets. 

All the above components require continuous field measurements. Attacks launched against any of these systems can potentially introduce delays and errors in time-critical operations. Without correct real-time information, the power grid is susceptible to human errors. Falsified data results in incorrect solutions to critical power system control algorithms.  This leads to increase in operation costs, incorrect generation dispatch, erroneous load shedding and even, widespread blackouts. Blackouts due to cyber attacks disrupt dependant water, gas and the internet networks, affecting a large number of customers and causing substantial monetary loss.

% The risks of large scale failures compound when  data attacks are carried out in coordination with other cyber and/or physical attacks. 

% The actual cost of power outage then increases at a much slower rate than the actual duration and is supported by the Weber-Fechner's Law which discusses the logarithmic relationship between the stimulus and the intensity of the impact \cite{Baarsma2009PricingNetherlands}. 

% This becomes significant as data falsification can make a control operator blind to the actual system.
% Without correct real-time information, the power grid grows susceptible to incorrect human decisions and subsequent failures such as loss of load or widespread blackouts. 

The rest of the paper is organized as follows. In Section \ref{sec:surveyofsurveys}, technical documents and surveys related to power system cyber-security are reviewed. Section \ref{sec:controlOperationBasics} provides an overview of various smart grid control operation blocks. Section \ref{sec:steadyState} outlines the consequences of attacks on steady state operations. Discussions on vulnerabilities affecting transient and auxiliary control are presented in Section \ref{sec:transientAux}. In Section \ref{sec:substation}, data attacks against substation architecture are highlighted while Section \ref{sec:AMI_markets} discusses consequences of energy alteration attacks. This is followed by a discussion section that 
highlights key similarities between different attack models, explores global cyber security standards and identifies future research directions. 

\section{Literature Survey}\label{sec:surveyofsurveys}

There has been much deliberation and investigation in areas of power systems and cyber security in the last couple of years. In this section, information from multiple official reports, guidelines and academic papers is concisely presented. Table \ref{tab:allReferences} compiles all major resources in this area of research.

In 2009, Sandia National Laboratories issued a report on cyber attacks against electric power systems \cite{Stamp2009SANDIAFY08}. This case study investigated attack impacts on the dynamic behavior of the grid using a finite state abstraction model. A three-volume report on implementation, design, operation, maintenance and mitigation of cyber-related issues was presented by NISTIR-7628 in 2010 \cite{TheSmartGridInteroperabilityPanel2010IntroductionSecurity}. In 2011, NERC published the requirements of critical infrastructure protection (CIP). This specifically aimed to highlight high, medium and low impact cyber threats on bulk electric system components such as control centres, substations, generation resources, remedial schemes and switching \cite{Huff-Arkansas2011CIPStatus}.  The Cyber Attack Task Force, in 2012, published a report on coordinated attacks that resulted in unnecessary circuit breaker operation, loss of load and generation, communication system disruption and cascading failures and proposed possible mitigation strategies \cite{NERCCyberReport}. The findings of this Task Force were influenced by a 2010 report  \cite{2010High-ImpactSystem}. In 2015, \cite{Campbell2015CybersecuritySystem} emphasized the need for additional financial and manpower investments to curb cyber vulnerabilities in bulk power systems. It looked into potential attackable areas in electric grids not covered under NERC CIP standards. Other key findings and reports include the Cyber Threat and Vulnerability Analysis that highlighted challenges to reliable operation of the grid and proposed best practices to utilities \cite{IdahoNationalLaboratory2017}. A 2018 report \cite{Schlichting2018AssessmentOSD} provides a broader perspective on threats against energy systems, highlights current practices for cyber-physical security and recommends best mitigation practices. 

In general, such best practices include developing and maintaining a secured network infrastructure, installation of firewalls and using updated antivirus software. Others include encryption of data, patch management, setting strong passwords, intrusion detection using deep packet inspection, storage of data for future analysis, use of secure application control software, purchasing equipment from reliable vendors, restricting physical access, employee training and organization wide cyber-security framework with an overall risk management plan.      

\begin{table*}[t]
\caption{Major Resources and References for Data Injection Attacks [2009 - 2019]}
\centering
\begin{tabular}{|c|c|}
\hline
\textbf{Resources and Areas} & \textbf{References}  \\ \hline
\makecell{Government and \\Not-for-profit organization reports} & \begin{tabular}[c]{@{}c@{}}\makecell{National Laboratories \cite{Stamp2009SANDIAFY08,IdahoNationalLaboratory2017}, NISTIR \cite{TheSmartGridInteroperabilityPanel2010IntroductionSecurity}, NERC\cite{Huff-Arkansas2011CIPStatus,NERCCyberReport}, DOE \cite{2010High-ImpactSystem}, \\ Congressional Research Service\cite{Campbell2015CybersecuritySystem}, MITRE Corporation\cite{Schlichting2018AssessmentOSD} }\end{tabular}   \\ \hline
Smart grid communication & \begin{tabular}[c]{@{}c@{}}\makecell{Data communication architecture\cite{Yan2013AChallenges,Ericsson2010CyberInfrastructure,Yan2012ACommunications,YilinMo2012CyberPhysicalInfrastructure,Mrabet2018Cyber-securityChallenges, Li2012SecuringChallenges,Jokar2016AGrids,Liu2012CyberGrids}}\end{tabular}  \\ \hline
Other Surveys & \begin{tabular}[c]{@{}c@{}}\makecell{Control loops \cite{Sridhar2012CyberPhysicalGrid}, Cyber-physical test beds \cite{Sun2018CyberState-of-the-art}, Security, safety and monitoring challenges \cite{Wang2013AGrid,Pour2017ASystems,Rasmussen2017AAssessment,Line2011CyberGrids, Kotut2016SurveyGrids}, State \\estimation,  AGC and electricity markets \cite{Chatterjee2017ReviewOperations,Deng2017,Liang2017}}\end{tabular}   \\ \hline
Steady State Control & \makecell{Static State Estimation \cite{Zhang2018CanSystems,Liu2009,QingyuYang2014OnCountermeasures,Hug2012,Liu2017FalseInformation, esmalifalak2011stealth,rahman2012false,Yu2015BlindGrid,Basumallik2017}, Dynamic State Estimation \cite{Chen2017AGrid},  Security Constrained Optimal Power Flow \cite{Khanna2017Bi-levelFlow},\\ Economic Dispatch \cite{Shelar2017CompromisingOperations},  Load Redistribution \cite{Yuan2011,yuan2012quantitative,Liu2014LocalInformation, Liu2015ModelingInformation,li2018analyzing,xiang2015power,xiang2015game,xiang2017framework,xiang2015coordinated,pinceti2018load}, Topology Falsification \cite{Liu2016MaskingAttacks,Kim2013OnCountermeasures,lin2013distributed}\\ Coordinated Attacks \cite{Li2016BilevelSystems, Deng2017CCPA:Grid}} \begin{tabular}[c]{@{}c@{}}\end{tabular} \\ \hline
Transient and Auxiliary Control &  \makecell{Rotor Angle Stability \cite{Farraj2017OnControl}, Automatic Generation Control\cite{Sridhar2014Model-BasedControl,Ashok2015ExperimentalTestbed,Tan2017ModelingControl},  Automatic Voltage/VAR Control \cite{Chen2018EvaluationControl,Teixeira2014SecurityCountermeasures} \\ FACTS devices \cite{Rubio-Marroquin2018ImpactParameters} }   \\ \hline
Substation Control & \makecell{PMUs \cite{Zhang2013,Pal2016,Varmaziari2017Cyber-attackObserver,Yuan2011SecurityCounters} IEC 61850 \cite{Rashid2014ANetwork,Kabir-Querrec2016ANetworks} Switching Attacks \cite{liu2014coordinated}}   \\ \hline
Energy Alteration Attack &  \makecell{AMI \cite{Li2017HMM-BasedInfrastructure,Liu2015AInfrastructure,Khanna2016DataProfit,Grochocki2012AMIRecommendations}, Residential Load Control \cite{Mishra2015RateGrid},  Distributed Energy Routing \cite{Lin2012OnGrid},  Micro-grids \cite{Zhang2015OnGrid,Chlela2016Real-timeAttacks,Chlela2018FallbackCyber-Attacks} \\Electricity Markets  \cite{Xie2010FalseMarkets,Lin2016TowardsMarkets}}   \\ \hline
\end{tabular}
\label{tab:allReferences}
\end{table*}

Apart from organizational findings and recommendations, a vast amount of research on cyber-security vulnerabilities and mitigation strategies have been published in academic journals. Authors in \cite{Yan2013AChallenges} have presented a thorough review of smart grid communication infrastructures and related cyber threats and challenges. Vulnerable access points in the SCADA/EMS architecture include: field staff remote access computers, local LAN switches, modems connecting Remote Terminal Units and SCADA/EMS Master, bridge/routers connecting to the Wide Area Network and station HMIs linked to protection relays \cite{Ericsson2010CyberInfrastructure}. Attacks exploiting these vulnerable access points include Distributed Denial of Service, reconnaissance, scanning, exploitation and access attacks that are often carried out through
malwares and viruses. Such attacks can spoof electricity prices, alter meter data and falsify control commands \cite{YilinMo2012CyberPhysicalInfrastructure,Mrabet2018Cyber-securityChallenges}. To minimize cyber attack impacts, the requirements for privacy, availability, integrity and authentication for SCADA communication networks are summarized in \cite{Yan2012ACommunications}. Encryption techniques, key management issues and security and privacy threats against smart meters, electric vehicles, distributed generations  are highlighted in \cite{Liu2012CyberGrids,Jokar2016AGrids,Rasmussen2017AAssessment,Line2011CyberGrids, Kotut2016SurveyGrids,Li2012SecuringChallenges}.

On the other hand, \cite{Sridhar2012CyberPhysicalGrid,Wang2013AGrid,Pour2017ASystems,Sun2018CyberState-of-the-art} discusses attacks on the physical side of the grid. A qualitative survey of cyber attacks on various control loops in power systems is presented in \cite{Sridhar2012CyberPhysicalGrid}. Attacks and impacts on real cyber-physical test beds are demonstrated in \cite{Sun2018CyberState-of-the-art} while bad data injections and their countermeasures are analyzed in \cite{Wang2013AGrid,Pour2017ASystems}.

\begin{figure*}[!htpb]\vspace{-21em}
\startchronology[startyear=2009,stopyear=2019, startdate=false, color=blue!40, stopdate=false, arrow=true, height=3pt]
\setupchronoevent{textstyle=\scriptsize,datestyle=\scriptsize}
\chronograduation[event]{100}
\chronoevent[markdepth=-60pt]{2009}{FDIA DC SE \cite{Liu2009}}

\chronoevent[markdepth=50pt]{2010}{FDIA Locational Marginal Price \cite{Xie2010FalseMarkets}}

\chronoevent[markdepth=-100pt]{2011}{PMU Code injection \cite{Yuan2011SecurityCounters}}
\chronoevent[markdepth=-60pt]{2011}{Load Redistribution Attack \cite{Yuan2011}}
\chronoevent[markdepth=-20pt]{2011}{FDIA DC SE incomplete information \cite{esmalifalak2011stealth}}

\chronoevent[markdepth=20pt]{2012}{FDIA DC SE incomplete information \cite{rahman2012false}}
\chronoevent[markdepth=60pt]{2012}{FDIA AC SE \cite{Hug2012}}
\chronoevent[markdepth=100pt]{2012}{FDIA Distributed Energy Routing \cite{Lin2012OnGrid}}

% \chronoevent[markdepth=20pt]{2012}{FDIA DC SE incomplete information \cite{rahman2012false}}
% \chronoevent[markdepth=60pt]{2012}{FDIA AC SE \cite{Hug2012}} 
% \chronoevent[markdepth=100pt]{2012}{FDIA Distributed Energy Routing \cite{Lin2012OnGrid}}

\chronoevent[markdepth=-30pt]{2013}{Topology Attacks \cite{Kim2013OnCountermeasures}}
\chronoevent[markdepth=-80pt]{2013}{PMU GPS Time Sync Attack \cite{Zhang2013}}

\chronoevent[markdepth=5pt]{2014}{Load Redistribution incomplete information \cite{Liu2014LocalInformation}}
\chronoevent[markdepth=45pt]{2014}{Attack against AGC \cite{Sridhar2014Model-BasedControl} }
\chronoevent[markdepth=75pt]{2014}{Data Attack volt/VAR control \cite{Teixeira2014SecurityCountermeasures}}
\chronoevent[markdepth=105pt]{2014}{IEC61850 Attacks \cite{Rashid2014ANetwork}}
\chronoevent[markdepth=140pt]{2014}{Switching Attacks \cite{liu2014coordinated}}

\chronoevent[markdepth=-15pt]{2015}{FDIA AMI\cite{Liu2015AInfrastructure}}
\chronoevent[markdepth=-53pt]{2015}{FDIA Residential Load Control \cite{Mishra2015RateGrid}}
\chronoevent[markdepth=-105pt]{2015}{FDIA Microgrid Partition \cite{Zhang2015OnGrid}}

\chronoevent[markdepth=25pt]{2016}{FDIA + Topology Attacks \cite{Liu2016MaskingAttacks}}
\chronoevent[markdepth=55pt]{2016}{Co-oridnated Attacks \cite{Li2016BilevelSystems}}
\chronoevent[markdepth=90pt]{2016}{PMU Packet Drop Attack \cite{Pal2016}}  
\chronoevent[markdepth=120pt]{2016}{FDIA Multistep Electricity price \cite{Lin2016TowardsMarkets}}

\chronoevent[markdepth=-10pt]{2017}{FDIA AC SE incomplete information \cite{Liu2017FalseInformation}}
\chronoevent[markdepth=-40pt]{2017}{FDIA Dynamic SE \cite{Chen2017AGrid,Karimipour2017OnGrids}}
\chronoevent[markdepth=-80pt]{2017}{FDIA Security Constrained OPF \cite{Khanna2017Bi-levelFlow}}
\chronoevent[markdepth=-120pt]{2017}{FDIA Economic Dispatch \cite{Shelar2017CompromisingOperations}}
\chronoevent[markdepth=-160pt]{2017}{FDIA Rotor Angle Stability \cite{Farraj2017OnControl}}
\chronoevent[markdepth=-190pt]{2017}{FDIA Automatic Generation Control \cite{Tan2017ModelingControl}}

\chronoevent[markdepth=10pt]{2018}{FDIA Automatic Voltage Control \cite{Chen2018EvaluationControl}}
\chronoevent[markdepth=50pt]{2018}{DoS Attack Mircogrid \cite{Chlela2018FallbackCyber-Attacks}}
\chronoevent[markdepth=80pt]{2018}{FDIA FACTS devices \cite{Rubio-Marroquin2018ImpactParameters}}

\chronoevent[markdepth=-60pt]{2019}{FDIA Microgrid Distributed Load Sharing \cite{Zhang2019DistributedMicrogrid}}
\stopchronology
\caption{\small Chronology of benchmark papers in cyber attack design and impact on power systems [2009-2019]}
	\label{fig:chronology}
	\vspace{-1em}
\end{figure*}

% In \cite{Rasmussen2017AAssessment}, broad challenges faced by the cyber-physical grid are highlighted in terms of security, safety and monitoring. 

% Security architectures of the smart grids, different attack scenarios and challenges faced by the electricity network are underlined in \cite{Line2011CyberGrids, Kotut2016SurveyGrids}. 

% A general taxonomy of various attacks targeting physical devices and networks is given in \cite{Li2012SecuringChallenges}. 

All the aforementioned resources delineate the growing cyber threats against energy grids. It thus  necessary to categorize the various scattered information into a single taxonomy of data attacks, shown in Fig. \ref{fig:taxonomy}. All attempts have been made to include as many major resources as possible. While authors in \cite{Chatterjee2017ReviewOperations,Deng2017,Liang2017} discussed quantitative attack models on state estimation, AGC and electricity markets, a large section of power system components vulnerable to attacks were overlooked. This paper presents a taxonomy that classifies nineteen different attack models into four broad areas, and hence, is more comprehensive. Besides, each attack scenario is supplemented with concise and necessary mathematical details, much of which have been omitted from previous surveys and reports in the literature. Additionally, the chronological development of cyber attack design is tracked, as presented in Fig. \ref{fig:chronology}. The overall intention of this paper is to highlight advanced cyber attack designs and their possible impacts on the grid. This can serve as a common platform to operators, researchers and stakeholders in both power system and cyber security domains.

\begin{figure*}[t]
	\centering
	\includegraphics[scale=0.39]{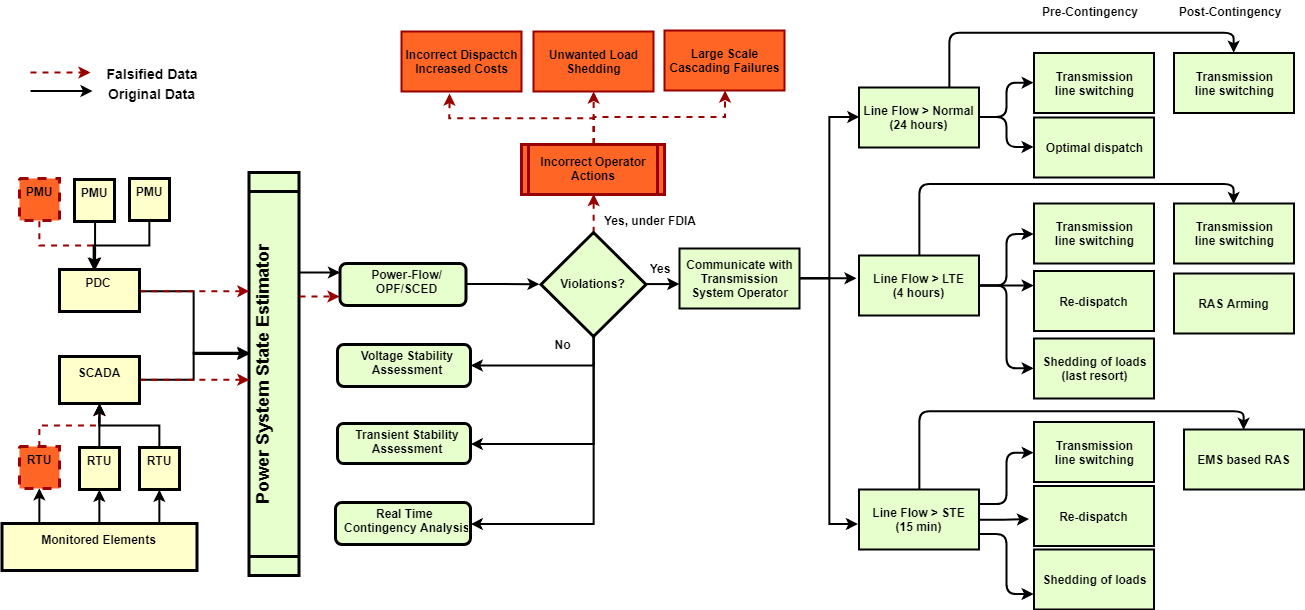}
	\caption{\small EMS architecture and operator actions under normal conditions and FDIA}
	\label{fig:RAS}
	\vspace{-1em}
\end{figure*}

\section{Control and Operation of Power System}\label{sec:controlOperationBasics} 

In this section, a brief review on key smart grid  control operations is presented. At the heart of the system is the master program called the Energy Management Systems (EMS), a high performance critical control system overlooking all monitoring, control and optimization functions. Inside the EMS is the state estimator (SE) program which collects redundant measurements from various SCADA and PMU devices, eliminates gross data, and estimates bus voltages and angles using a non-linear weighted least squares method \cite{Schweppe1970PowerModel}. SE provides a  quasi-static model of the power system under real-time operating conditions using PMU and traditional measurements. The output of SE is used as the basis for the critical applications in EMS including power-flow and economic dispatch, load forecast, voltage security, transient and small signal stability, real-time contingency analysis (RTCA) and steady state remedial action schemes (RAS). 

Other services and control blocks include (a) AGC responsible for maintaining system wide frequency by regulating power output of generators and reducing area control error~\cite{Jaleeli1992UnderstandingControl}, (b) maintaining generation reserves to ensure reliability in the event of generation loss and (c) performing transfer and contingency analysis. Transfer analysis helps determine the extent by which the current operating system can be moved before being bounded by security limit violations. Once transfers between different areas are identified, system operators perform $N-1$ or $N-1-1$ contingency analysis to secure the system under any real time operating conditions \cite{2007NERCVer.:l.0.2}. After a contingency has occurred, transient stability program plays a central role in regulating synchronous generator load angles within acceptable limits at a short period of time \cite{Kundur1994PowerControl}. Ancillary services help preserve transmission balance by maintaining spinning and non-spinning reserves and managing load regulation and voltage support \cite{Rebours2007AFeatures,Pirbazari2010AncillaryWorld}. Volt-Var compensation techniques ensure that the voltages and reactive power remain within their desired limits during both normal or emergency conditions \cite{Grainger1985Volt/VarProblem}. Fault location, isolation, and service restoration (FLISR) is another important control block used to analyze, isolate faults and help restore services \cite{Parikh2013FaultGOOSE,Habib2017Multi-Agent-BasedRestoration}. FLISR often employs time synchronized phasor measurements for a more accurate estimation of fault location. To safeguard the system against cascading failure after a major fault, controlled islanding techniques partition the system into smaller self-sufficient micro-grids with minimum load generation imbalance. For more accurate real-time analysis of the grid, utilities often deploy redundant measurement devices. Such redundant  systems include multiple voltage, current, injection and flow measurements from different substations across the network, redundant generations, parallel substation communication architecture, backup protective relays, SCADA computers, logs and databases \cite{Castello2015ABus,Antonova2011CommunicationAutomation}. 

All the above sensors, which monitor real time measurements and are responsible for time-critical operations, are susceptible to cyber attacks. Data falsification can result in inaccurate solutions for different control blocks, leading to incorrect operator decisions, uneconomic dispatch, load shedding or large scale power outages. The rest of the paper focuses on the mathematical modeling and impact assessment of data falsification attacks against four control operation blocks: (a) steady state control, (b) transient and auxiliary control, (c) substation control and (d) energy/load control. 

\section{Attacks on Steady State Control}\label{sec:steadyState} 
In this section, cyber attacks against various steady state control blocks are considered. These blocks include state estimation, optimal power flow and economic dispatch. The structure of EMS and associated operator actions are shown in Fig.~\ref{fig:RAS}. Attacks involve data and topology falsification at the device level with the objective to alter the solution of the control algorithms. Incorrect solutions can induce line overloads that can lead to unwanted tripping, increase in generation costs and in the worst case, extensive power failure.  

\subsection{Static State Estimation}

Static state estimation provides quasi-real-time voltage magnitude and phase angle at each bus of the network under real-time operating conditions. The states are estimated by solving an over-determined solution of non-linear power flow equations using measurements obtained from traditional SCADA meters and PMUs~\cite{Monticelli2000}.

General assumptions to carry out successful attacks include (1) availability of complete or partial system topology and Jacobian matrix, (2) non-attackable zero-injection buses, (3) non-attackable generator measurements or attackable small distributed generators, (4) access to dynamic line limit monitoring devices, meter measurements and meter ID mapping in EMS and (5) available historical load, generation capacity, cost, dispatch sequences and locational marginal price data~\cite{Deng2017CCPA:Grid,Zhang2018CanSystems,Liu2009,QingyuYang2014OnCountermeasures,Hug2012,esmalifalak2011stealth,rahman2012false,Yu2015BlindGrid,Liu2014LocalInformation,Shelar2017CompromisingOperations,Basumallik2017}.

Attack against linear state estimation was first demonstrated by Liu \textit{et al.} \cite{Liu2009}. Yang \textit{et al.} extended \cite{Liu2009} to find sparsest attack vectors that caused maximum deviation in the estimated states. To find such sparsest attack vectors, linear transformation of Jacobian matrix and heuristic approaches were employed \cite{QingyuYang2014OnCountermeasures}. On the other hand, FDIA against non-linear AC state estimator were investigated by Hug \textit{et al.} in \cite{Hug2012}. With complete system information available to the attackers, authors in \cite{Hug2012} proposed to inject attack vectors such that (a) the injected attack vector satisfy KCL at each node and (b) the residual of the altered SE is below the set threshold,

\begin{equation}
\begin{aligned}
\begin{Vmatrix}
z_{bad}-h(\hat{x}_{2bad})
\end{Vmatrix}
& = \begin{Vmatrix}
\begin{pmatrix}
z_1\\ 
z_{2bad}
\end{pmatrix}- 
\begin{pmatrix}
h(\hat{x}_1)\\ 
h(\hat{x}_{2bad})
\end{pmatrix}
\end{Vmatrix}\\
& = 
\begin{Vmatrix}
\begin{pmatrix}
z_1\\ 
z_2+a
\end{pmatrix}- 
\begin{pmatrix}
h(\hat{x}_1)\\ 
h(\hat{x}_{2}+c)
\end{pmatrix}
\end{Vmatrix}\\
& =
\begin{Vmatrix}
\begin{pmatrix}
z_1 \\ z_2
\end{pmatrix}
-\begin{pmatrix}
h(\hat{x}_1)\\ 
h(\hat{x}_2)
\end{pmatrix}
\end{Vmatrix}
< \tau 
\end{aligned}
\end{equation}
Here, the measurement set $z$ divided into either those that are compromised ($z_2$) or those that are not ($z_1$), $a$ is the attack vector, $c$ is the bias added to the original states and $h$ is the set of non-linear power flow equations relating the states to the measurements. The attack vector can then be constructed as $a=h(\hat{x}_2+c)-h(\hat{x}_2)$. 

The assumption that complete network information is available with attackers was challenged by authors in \cite{Liu2017FalseInformation, esmalifalak2011stealth, rahman2012false, Yu2015BlindGrid}. The feasibility of FDIA under incomplete information of circuit breaker status, transformer taps, network connectivity matrix and admittance matrix was established by Rahman \textit{et al.}~\cite{rahman2012false}. In case of partial network information, authors in~\cite{Liu2017FalseInformation} proposed the following optimization problem to construct FDIA vectors,
\vspace{-1em}
\begin{equation}
\begin{aligned}\label{fdia}
& {\text{minimize}}
& & \sum_{i=1}^{10} 1^T S_i\\
& \text{subject to} & &  \begin{bmatrix}
\Delta P\\ 
\Delta Q\\ 
\Delta p\\ 
\Delta q\\ 
\Delta V\\ 
\end{bmatrix} =\begin{bmatrix}
\partial P/\partial V & \partial P/\partial \delta\\ 
\partial Q/\partial V & \partial Q/\partial \delta\\ 
\partial p/\partial V & \partial p/\partial \delta\\ 
\partial q/\partial V & \partial q/\partial \delta\\ 
1 & 0
\end{bmatrix} \begin{bmatrix}
\Delta V\\ 
\Delta \theta
\end{bmatrix}\\
& & & P^{min} \leq P+\Delta P+S_1-S_2 \leq P^{max}\\ & & & p^{min} \leq p+\Delta p+S_3-S_4 \leq p^{max}\\
& & & q^{min} \leq q+\Delta q+S_5-S_6 \leq q^{max}\\
& & & V^{min} \leq V+\Delta V+S_7-S_8 \leq V^{max}\\
& & & \delta ^{min} \leq G(\delta+\Delta \delta)+S_9-S_{10} \leq \delta^{max}\\
\end{aligned}\\
\end{equation}

\noindent Here,  $P,Q,p,q$ are bus active and reactive power injections and line flows respectively, $V$ is bus voltages, $G$ is the transition matrix transforming bus voltage angles $\theta$ into line angle differences and $S_i$ are slack variables. The new attack vector can be updated as $[V \;  \delta]^T=[V \;  \delta]^T+[\Delta V \; \Delta  \delta]^T$. Other approaches include those by Esmalifalak \textit{et al.} \cite{esmalifalak2011stealth} where topology and states were inferred from power flow measurements using independent component analysis.  General blind FDIA was introduced by Yu and Chin \cite{Yu2015BlindGrid} where an approximation of Jacobian matrix was obtained using principal component analysis. Such approximated attack vectors were proved to be almost stealthy. 

With many time-critical control blocks in power systems depending on the SE output, a direct attack targeting the SE solution will interfere with the real time operation and control of the grid. Data falsification will lead to incorrect state estimation, mislead power flow solutions, resulting in incorrect line tripping, unwanted re-dispatch of generation, or in the worst case, load shedding and cascades.

\subsection{Dynamic State Estimation}
There are two major drawbacks of the traditional state estimator: (1) it fails to capture the dynamics of the power system due to continuous random load fluctuations and generation adjustments, (2) slow convergence and large computation power is involved in estimating states at short intervals for large networks. The above two drawbacks are addressed by the dynamic SE \cite{Debs1970ASystem}. Dynamic SE are faster and are able to predict the future states at $t+1$ time using the knowledge of the system at $t^{th}$ time step. The process of dynamic SE are summarized through the two equations,  $x_{t+1}=F(t)x(t)+G(t)+w(t)$ and $z_{t+1}=h(x(t+1))+e(t+1)$. Here, $x$ refers to the estimated states, $F$ is the state transition matrix, $G$ is the state trajectory vector, $w$ is the Gaussian noise added to the signal, $z$ is the set of measurements linking the non-linear power flow equations $h$ to the states and $e$ is Gaussian measurement noise. Authors in \cite{Chen2017AGrid,Karimipour2017OnGrids} have investigated attacks against  dynamic SE. A successful attack vector $a$ was constructed as, 
\begin{equation}
    ||a(k + 1) + h(\hat x(k+1))-h(\hat x_a(k+1))|| \leq \epsilon
\end{equation}{}

\noindent where $\tau$ is the maximum tolerance of error of the estimator and $0 \leq \epsilon \leq \tau$. Such  false data injection introduces bias in specific states, which can further corrupt power flow and other control algorithms similar to attack on traditional SE.

\subsection{Security Constrained Optimal Power Flow}
After estimating the states, optimal power flow is solved to find transmission line power flows. This is done by power flow algorithms which solve a set of non-linear power balance (generation, load and network) equations \cite{Kundur1994PowerControl}. An extension of the optimal power flow solution is the security constrained optimum power flow (SCOPF). SCOPF has added constraints for generator power limits, transmission line capacity and contingency constraints that ensure the system is both pre-contingency and post-contingency stable with no limit violations \cite{Capitanescu2007ContingencyFlow,Wibowo2014SecurityControl,FengDong2012PracticalFlow}.

Falsified state estimates result in incorrect SCOPF solutions, jeopardizing real-time reliability of the grid. A direct attack on the SCOPF algorithm for a DC powerflow model was discussed by Khanna \textit{et al.} \cite{Khanna2017Bi-levelFlow}. A bi-level optimization problem was formulated where an attacker falsified load measurements by compromising minimum number of meters. The objective was to alter generation dispatch such that the system becomes susceptible to a single point failure. The outer optimization problem was formulated as, \begin{equation}
\begin{aligned}\label{loadDistribute}
& \underset{{\Delta P_D}}{\text{Min}}
& & {\sum_{\forall d \in N_D}L_d+2 \sum_{\forall l \in N_L}FL_l}\\
\end{aligned}
\end{equation} 
Here, $L_d$ is a binary variable which is one if there is a change in the load measurement at bus $d \in N_D$, and $FL_l$ is another binary variable indicating changes in the flow measurement $\Delta P_D$ of transmission line $l \in N_L$. The objective minimizes the number of meters compromised with the following constraints: (a) changes in measurements were confined within tolerable limits, (b) net load in the system was kept unchanged, (c) altered line flows were below line limits and (d) that the new dispatch was not $N-1$ compliant. The system failure was guaranteed by ensuring that there exists line overloads following a single line disconnection with the new dispatch. The inner level of the optimization is the SCOPF with the perturbed measurements to obtain the new dispatch $P_g$,

\begin{equation}
    P_g = \text{arg} \thinspace \{{ \text{min}_{P_g}}\sum_{i \in N_G}C_{gi}P_{gi}\}
\end{equation}
where $C_{gi}$ is the cost associated with dispatch $P_{gi}$. The outer level optimization was solved through meta-heuristic techniques while quadrature programming was employed to solve the inner level problem. Feasible attacks led to mis-operation, resulting in multiple line trips following a single line outage. Attacks against SCOPF also result in misleading Automatic Voltage Regulators (AVC), details of which are described in Section \ref{sec:transientAux}-C.

\subsection{Economic Dispatch}

The economic dispatch algorithm is solved periodically by system operators to balance load and demand at the lowest operating cost, subjected to transmission and generation capacity constraints \cite{Abido2006MultiobjectiveProblem,Gaing2003ParticleConstraints}. The output of economic dispatch is heavily dependant on the SE solutions \cite{Chowdhury1990ADispatch}. 

Authors in \cite{Shelar2017CompromisingOperations} propose a linear-quadratic bi-level optimization considering data attacks on the economic dispatch algorithm. The main aim of the attacker is to alter power flows and congest critical lines in the network. The attackers first adjust the dynamic line ratings (DLR) of transmission lines (equipped with dynamic line limit monitoring devices) to maximize line flow $f_{ij}$ violations,

\begin{equation}
\begin{aligned}
& \underset{{i,j} \in E_D}{\text{Max}}
& & 100\left ( \frac{|f_{ij}|}{u_{i,j}^d}-1 \right )\\
\end{aligned}
\end{equation} 

\noindent Here, $u_{i,j}^d$ and $u^a$  are the correct and falsified DLR. Assumptions include attackers ability to alter dynamic line ratings, prior knowledge of system topology, nodal power injections, generator capacities, transmission line limits and generation costs. On the other hand, the operator minimizes operation cost $C_i$ with the falsified measurements for a dispatch $p_i$,
\begin{equation}
\begin{aligned}
&  \text{Min}& &  C(p)=\sum_{i \in G} C_i(p_i)\\
\end{aligned}
\end{equation} 
The optimization problem was subjected to power flow constraints, load-generation constraints and flow capacity constraints. The linear-quadratic bilevel optimization was reformulated as a Mixed Integer Linear Program (MILP) using the Karush-Kuhn-Tucker (KKT) conditions for the lower level problem. Crafted attack against the economic dispatch algorithm can result in increase operation costs, line overloads and subsequent line tripping and further aggravating the risks of cascading failures. 

An extension of the economic dispatch problem is the  security constrained economic dispatch (SCED) \cite{FederalEnergyRegulatoryCommission2006SecurityRecommendations,Monticelli1987Security-ConstrainedRescheduling}. SCED minimizes operating costs to meet customer demands while ensuring transmission line capacities are not violated following any contingency \cite{Vargas1993ADispatch,Elacqua1982SecurityPool}. Authors in \cite{Yuan2011} have formulated attacks against SCED as load redistribution attacks that modify load measurements at different buses, which are considered next.

\subsection{Load Redistribution Attacks}

Load Redistribution (LR) attack is a class of false data injection attacks that aims to compromise the SE and SCED/SCOPF by falsifying load measurements \cite{Yuan2011,yuan2012quantitative,Liu2014LocalInformation, Liu2015ModelingInformation,li2018analyzing,xiang2015power,xiang2015game,xiang2017framework,xiang2015coordinated,pinceti2018load}. In LR attacks, load injections measurements are increased at certain buses and decreased at other buses keeping total load unchanged and altering corresponding power flow measurements so that KCL holds at every node. 

Yuan \textit{et al.} identified  FDIA as LR attacks \cite{Yuan2011}. Such attacks were carried out by changing load measurements up to 50\% of their original set point, leading to a secure operating condition with higher operation costs. The LR attack problem was formulated as a bi-level optimization problem. With the total generation cost as $c_gP_g$, load shedding cost as $cs_dS_d$, load shedding amount as $S_d$ altered load as $\Delta D$, load alteration tolerance as $\tau$, shift factor matrix as $SF$ and bus load incidence matrix as $KD$, the upper level problem aimed to maximize generation and load-shedding costs,

\begin{equation}
\begin{aligned}\label{distribute}
& max_{\Delta D}
& & \sum_{g=1}^{Ng}c_gP_g^*+\sum_{d=1}^{Nd}cs_dS_d^*\\
& s.t. & &  \sum_{d=1}^{Nd} \Delta D_d = 0\\
& & &  -\tau D_d \leq \Delta D_d \leq \tau D_d \\
& & &  \Delta PL = -SF. KD. \Delta D 
\end{aligned}
\end{equation} 

\noindent The optimization problem was subjected to other logical, resource and SCED constraints. It also ensured that total load demand remained unchanged and that load measurements were varied within specific tolerance. The system operator on the other hand minimizes operation cost by re-dispatching generation or by shedding loads. The operator actions are based on falsified states obtained after an incorrect SE solution. The SCED for the operator's side was formulated as,
\begin{equation}
\begin{aligned}\label{loadDistribute5}
\{P_g, S_d\} = &\thinspace \text{arg} \{\thinspace {\text{Min}}
& & \sum_{g=1}^{N_g}c_gP_{g}+\sum_{d=1}^{N_d}cs_dS_{d}\}\\
%&  \text{s.t.}& &  \sum_{g=1}^{N_g}P_{g}=\sum_{d=1}^{N_d}(D_d - S_d)\\
\end{aligned}
\end{equation}
The authors transformed the bi-level problem to an equivalent single level problem using the Karush-Kuhn-Tucker (KKT) conditions to obtain the optimal attack vector.

Authors in \cite{Yuan2011ModelingSystems,yuan2012quantitative} considered altered load measurements not to exceed 50\% of the original value and maximum compromised meters were limited to 20. The LR attack transferred loads from multiple buses to the largest load bus in the system. In \cite{li2018analyzing}, authors considered maximum 20\% changes in the load measurement. Authors in \cite{xiang2015power} carried LR attacks on 10, 15, 25, and 30 meters and load changes were subjected to normal distribution with $N(20, 100)$. Smaller set of attackable measurements reflect limited attacker budget, thus more practical. Additionally, feasibility of LR attack under partial network information was considered by Liu \textit{et al.} in  \cite{Liu2014LocalInformation, Liu2015ModelingInformation}. 

LR attacks were shown to increase operating costs by (a) driving the SCED/SCOPF to uneconomic operating conditions resulting in non-optimal generation dispatch, (b) causing load shedding and (c) initiating line overloads and tripping. If such attacks are not discovered quickly, multiple overloaded line trips can initiate widespread load-shedding in the absence of immediate corrective actions. 

\subsection{Topology Falsification Attacks}
For reliable operation of the power system, it is crucial for utilities to quickly identify outaged lines.  
This is done by topology processor that estimates current system topology from circuit breaker contact signals hardwired to Remote Terminal Units (RTUs) prior to each cycle of SE. Under normal condition, authors in \cite{Liu2016MaskingAttacks} explain that line outage residuals are good indices to identify disconnected lines. When $k^{th}$ line is out of service, the residual is given as, $r_k=min_{f_k^0} ||\Delta\theta_{m,k}-\Delta\theta_{m,k}^{cal}||_2$. Here, $\Delta\theta_{m,k}$ and $\Delta\theta_{m,k}^{cal}$ are the observed and calculated phase angle changes observed by PMU. 

In \cite{Liu2016MaskingAttacks}, it was demonstrated that line outage residuals are potential candidates for successful attacks. After altering the line open status from 0 to 1 in the PMU packet data, the attackers need to falsify injection measurements at both end of the candidate transmission line. Two conditions are needed to lauch successful attacks: (1) the line outage residuals of the attacked line should be above a minimum threshold and (2) attacks should satisfy both KCL and KVL to successfully pass the existing bad data detection system in the SE. The ultimate goal of such attacks is to mask line outages by sending a combination of incorrect status and altered measurements to the control center.  A bi-level optimization problem was proposed where in the upper level problem, the attacker maximizes the candidate line residual above the minimum threshold,
\begin{equation}
\begin{aligned}
& {\text{Max}}
& & min_{f_k^0} ||\Delta\theta_{m,k}-\Delta\theta_{m,k}^{cal}||\\
\end{aligned}
\end{equation}
The objective function is constrained to load injection and power flow measurement alteration within tolerable ranges while ensuring that the total injected load remains unchanged. For the lower level problem, the operator aims to minimize the residuals,
\begin{equation}
\begin{aligned}
& {\text{Min}}
& & min_{f_k^0} ||\Delta\theta_{m,k}-\Delta\theta_{m,k}^{cal}||\\
\end{aligned}
\end{equation}
When the attacker modifies power injection and line flow measurements, the value of $\Delta\theta_{m,k}^{cal}$ changes to $\Delta\theta_{m,k}^{'cal}$. The above bi-level problem was re-formulated as a single level problem. The single level problem was non-convex and was written as two linear programs. Load injection measurement variations were kept within 50\% the rated values. Other approaches to topology attacks include falsifying of circuit breaker status \cite{Kim2013OnCountermeasures,lin2013distributed}.

The consequences of topology attacks can be understood straightforward: when line outages are masked and system topology is incorrect, observability analysis, state estimation and bad data detection fails resulting in erroneous line flows. This directly affects critical control programs like SCED, SCOPF and RTCA.

\subsection{Coordinated Attacks}

\begin{figure}[t]
	\centering
	\includegraphics[scale=0.31]{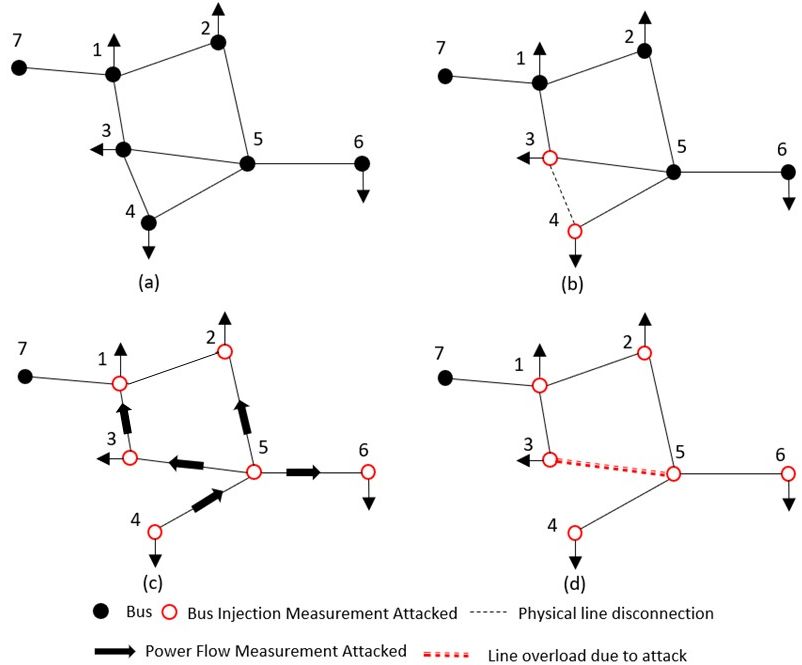}
	\caption{\small Coordinated attacks and consequences: (a) healthy power system, (b) line 3-4 physical outage coordinated with alteration of node injection measurements at 3 and 4, (c) falsification of power flow and injection measurements in attack neighborhood surrounded by injection buses, resulting in (d) line 3-5 physical overload.}
	\label{fig:cca}
	\vspace{-1em}
\end{figure}

Coordinated attacks on the other hand are the worst kind of attacks that combine FDIA, LR and topology falsification, as shown in Fig. \ref{fig:cca}. These attacks were considered by authors in \cite{Li2016BilevelSystems, Deng2017CCPA:Grid}. Li \textit{et al.} formulated a bi-level optimization problem where coordinated attacks were launched to mask line outages \cite{Li2016BilevelSystems}. Deng \textit{et al.} in \cite{Deng2017CCPA:Grid} combined physical attacks (designed as altered power flow measurements) with carefully constructed false data injection. 

A bi-level mixed-integer program was proposed in \cite{Li2016BilevelSystems}. Here, the upper level optimization maximized the impacts of physical attacks while the lower level optimization minimized attack costs by finding an optimal set of measurements to be altered,
\begin{equation}
\begin{aligned}
    \text{Upper Level: }\textbf{maximize}  \sum_{l \in L}[|\Delta PL_l^{(1)}|.(1-w_l)] \end{aligned} \end{equation}
\begin{equation}\begin{aligned}
\text{Lower Level: }\textbf{minimize}  \sum_{i \in B}c_{B,i}.\delta_{B,i}+\sum_{l \in L}c_{L,l}.\delta_{L,l} \end{aligned}
\end{equation} Here, $\Delta PL_l^{(1)}$ is the change in line flows after physical attack, $c_{B,i},c_{L,l}$ are resources needed to alter bus and line measurements and $w_l, \delta_{B,i},\delta_{L,l}$ are binary variables.
The optimization problem was subjected to budget, connectivity, topology, load redistribution, line limits and logical constraints \cite{Li2016BilevelSystems}. The mixed-integer nonlinear program was reformulated as a mixed-integer linear program. However, due to non-convexity of the lower level problem, KKT or duality based methods could not be used, and a two-stage sequential approach was proposed. Coordinated attacks were shown to result in multiple line overloads, unintended islanding and cascading failures.

\section{Attacks on Transient and Auxiliary Control}\label{sec:transientAux} 
In addition to steady state algorithms, transient and auxiliary control algorithms can also become subjected to targeted data attacks. These control algorithms oversee both local and global stability of the system under transient disturbances. Crafted attacks against these transient control blocks result in generator speed deviations leading to global synchronism loss and transmission line trips, thereby driving the system into states of emergency. Attacks against auxiliary controls can inhibit power transfer capability, interfere with voltage limit and power quality control.   

\subsection{Rotor Angle Stability}

Transient stability monitoring is a critical real-time power system control block that ensures synchronous generator load angles return to a steady state within a few cycles following a disturbance \cite{Kundur1994PowerControl,Guo2001GlobalSystems}. This is done by various external control mechanisms that are capable of controlling injected power in a short period of time. One such system is a decentralized external storage which is responsible for quickly injecting/absorbing power to/from the network after a fault. 

Authors in \cite{Farraj2017OnControl} investigated false data attacks against such storage devices and studied their impacts on rotor speed and angle. Data attacks against rotor speed of the generator w.r.t synchronous speed $\theta$, accelerating power of generator $P_A$ and rotor angle $\delta$ are formulated as,
\begin{equation}
\begin{aligned}
& \hat \theta_i (t) =  \theta_i (t) + f_{\theta i}(t)\\
&\hat P_{Ai} (t) =  P_{Ai} (t) + f_{P i}(t)\\
&\hat \delta_{i} (t) =   \delta_{i} (t) + f_{\delta i}(t)
\end{aligned}
\end{equation} where the attack signals are defined as, 
\begin{equation}
\begin{aligned}
& f_{\theta i}(t)= \epsilon_{i1}\theta_i (t)+\kappa_{i1}(t)+\mu_{i1} \\
& f_{P i}(t)= \epsilon_{i2}P_{Ai} (t)+\kappa_{i2}(t)+\mu_{i2}\\
&f_{\delta i}(t)= \epsilon_{i3}\delta _i (t)+\kappa_{i3}(t)+\mu_{i3}
\end{aligned}  
\end{equation}
Here $\epsilon, \kappa ,\mu$ are attack variables which control magnification, additive and bias components in the attack signal. When data is forged, the transient stability control may fail to maintain balance between electrical and mechanical torques after a major fault. This will cause rotor angle and speed to deviate from the desired set points, subsequently disconnecting generators from the grid and introducing shocks to the system.

% The major impact of such an attack is that the transient stability algorithm results in incorrect control outputs leading to increase in magnitude of rotor angle or divergence of rotor speed and angle with time.

\subsection{Automatic Generation Control}

Automatic generation control is another fundamental control block in power system whose objective is to maintain a nominal grid frequency \cite{Jaleeli1992UnderstandingControl,Apostolopoulou2014AutomaticTime}. The IEEE standard \cite{1970IEEESystems} defines the function of AGC as, ``the regulation of the power output of electric generators within a prescribed area in response to changes in system frequency, tie-line loading, or the relation of these to each other." 

Investigations in \cite{Sridhar2014Model-BasedControl} have demonstrated attack impacts on AGC due to signal magnifications and random packet delays. With the original measurement set $y(t)$, a scaling parameter $\lambda_s$ was used to alter the measurements,
\begin{equation}
    y'(t) = (1+\lambda_s)*y(t)
\end{equation}
In another case, a ramping parameter $\lambda_r$ was used to gradually alter the measurements as,
\begin{equation}
        y'(t) = y(t) + \lambda_r * t
\end{equation}
In both the cases, the corrupt measurements indicated an excess of power generation in one area. This lead to an incorrect AGC action to ramp down generation, eventually leading to frequency decay and load shedding. In \cite{Ashok2015ExperimentalTestbed}, a cyber security test bed was used to demonstrate tie-line and frequency measurement alteration that led to increase in area control error.

In \cite{Tan2017ModelingControl}, authors have illustrated data injection attacks targeted at AGC through a closed-form Laplace domain model. It was shown how an attacker was able to snoop on sensor data, gather knowledge about generation inertia and load damping parameters, and then use linear regression-based learning techniques to extract the required attack model. The input-output relationship of the AGC in terms of load change $\Delta p$, attack vector $a$ and frequency deviation $\Delta w$ is given by,
\begin{equation}\label{eq:AGC}
\hat{\Delta w}=\theta^T\Phi^{-1}\hat{\Delta p}+\theta^T\Phi^{-1}\Lambda\Psi T \hat{a}
\end{equation}
where $\Lambda$ and $\theta^T\Phi^{-1}$  are the second and fourth order polynomial fractions of the Laplace variable, $T$ is a matrix aggregating real-tie line flows into virtual tie-lines and $\Phi$ represents generator transfer function expressions. Based on eq (\ref{eq:AGC}), the attack vector in the $k^{th}$ AGC cycle was proposed as,
\begin{equation}
\Delta w(k) = \sum_{h=0}^{H-1} u_h^T \Delta p_{k-h}+v_h^T T a_{k-h}
\end{equation}
where $H$ is the regression horizon and $u$ and $h$ represent $\theta^T\Phi^{-1}$ and $\theta^T\Phi^{-1}\Lambda\Psi $ respectively. Detailed description on all variables can be found in \cite{Tan2017ModelingControl}. 

Sequential attacks targeting the AGC result in frequency deviation from the nominal set point. This can lead to mis-operation of remedial action schemes, disconnect generators and loads and damage expensive equipment, eventually triggering cascading failures. 

\begin{figure*}[t]
	\centering
	\includegraphics[scale=0.51]{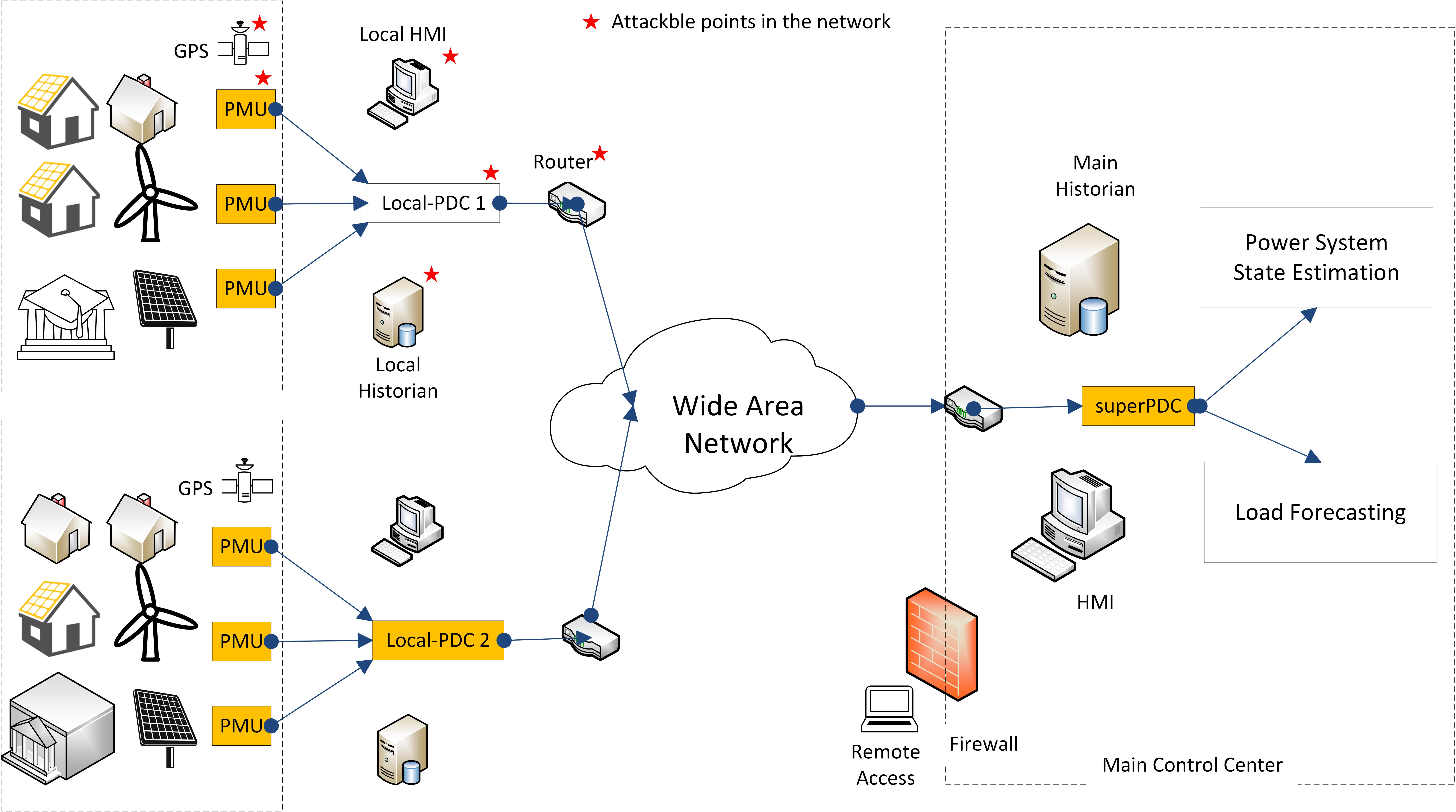}
	\caption{\small Common local vulnerable points in a PMU-PDC network}
	\label{fig:pmuAttack}
	\vspace{-1em}
\end{figure*}

\subsection{Automatic Voltage Control and Volt-VAR Control}

Automatic Voltage Control block maintains system voltages within desired range through reactive power injections and is commonly used to monitor inter-area reactive power balances \cite{Rebours2007AFeatures,Yoshida2000AAssessment}. AVC is highly dependent on the results of the SE. The operation of AVC can be briefly explained as follows: first, the output of the SE (voltages and angles) are fed as an input to the OPF. Once the OPF converges to a valid solution, the outputs are used to issue commands to generators triggering them to vary their output reactive power to maintain voltage within its prescribed margin. AVC aims to change active and reactive power by minimizing line losses and can be formulated as, \begin{equation}
\begin{aligned}\label{voltage}
&  {c(t)=\text{arg min } \{ \sum P^G(t) - \sum P^L(t)}\}
\end{aligned}
\end{equation} 
where $P^G(t)$ and $P^L(t)$ are active power generation and consumption respectively and $c(t)$ denotes optimal regulations of power generations \cite{Chen2018EvaluationControl}. The above optimization problem is subjected to all physical power system constraints. 

Authors in \cite{Chen2018EvaluationControl} investigated how falsified data can distort the functioning of the AVC leading to bus voltage violations at compromised substations. Such attacks send erroneous active and reactive power information to the dispatch centers triggering incorrect generator actions. The problem was formulated as a  Partial Observable Markov Decision Process (POMDP). The process of constructing a successful attack against AVC can be described as follows: first, the attacker changes active and reactive power measurements from $z$ to $\hat z$, which results in incorrect state estimation. The incorrect solution is fed to the OPF program whose outputs trigger changes in the voltage regulation commands.  When carried out during peak load hours can adversely affect closed-loop control systems required to maintain voltage stability and power balance. In the worst case, falsified measurements can also lead to system-wide outages. Further investigations on attacks at both ends of the line in a substation resulted in a total voltage collapse.

Volt-VAR control is another important control block ensuring that voltages at each point along the distribution feeder is maintained within the acceptable range of 0.95 p.u. to 1.05 p.u \cite{Grainger1985Volt/VarProblem,Roytelman1995Volt/varSystem}. Transformer load tap changers are used to maintain the voltage level while capacitor banks are used to control the reactive power. In a distribution network, consumer current loads $I_{ij}$ (denoted as the unknown state $y$) and feeder voltages $v_0$ are modeled as,
\begin{equation}
\begin{bmatrix}
v^k - v_0^k f_1(C_k)\\
(S^k/v_0^k)^*-v_0^kf_2(C_k)
\end{bmatrix}=\begin{bmatrix}
H_v(C_k)\\ 
H_S(C_k)
\end{bmatrix}y
\end{equation} 
where $S$ is the complex power, $f_i$ are scaling for capacitor bank arrangement $C$ and $H_v$ and $H_S$ are obtained from KVL equations  \cite{Teixeira2014SecurityCountermeasures}.

Authors in \cite{Teixeira2014SecurityCountermeasures} studied the feasibility of such attacks under different capacitor bank configurations. False data injected caused the Volt/VAR controls to issue incorrect commands to load tap changes and capacitor banks, which in turn increased the voltage level from $v$ to $v+a$. The new state description becomes,\begin{equation}
\begin{bmatrix}
v^k+a - v_0^k f_1(C_k)\\
(S^k/v_0^k)^*-v_0^kf_2(C_k)
\end{bmatrix}=\begin{bmatrix}
H_v(C_k)\\ 
H_S(C_k)
\end{bmatrix}(y+\Delta y)
\end{equation} 
It was found that for a successful attack under all possible capacitor configurations, the attack vector can be constructed as $a=H_v(C_1)B_c \alpha$ resulting in state bias $\Delta y = B_c \alpha$. Falsified data increase system operating costs and disrupt the normal operations of Volt/VAR control by making it unable to bring back voltages to desired set points.

\subsection{FACTS Devices}

FACTS devices are power electronic based solutions used to improve the controllability and power transfer capability in power systems \cite{1997ProposedFACTS}. These thyristor controlled devices are used to to manage active and reactive power and have a superior performance over traditional shunt capacitors and reactors. For a network, the terminal power measurements can be written as \cite{Rubio-Marroquin2018ImpactParameters},

\begin{equation}
    P_k = \frac{-V_kV_m sin(\theta_k-\theta_m)}{X_{TCSC}}
\end{equation}
\begin{equation}
    Q_k = \frac{-V_k^2+V_kV_mcos(\theta_k-\theta_m)}{X_{TCSC}}
\end{equation}
where $X_{TCSC}$ is the equivalent reactance of the TCSC, $P,Q,V, \theta$ are active, reactive power, bus voltage and angle respectively. 

Authors in \cite{Rubio-Marroquin2018ImpactParameters} have studied the impacts of false data injection on two FACTS devices: thyristor controlled series compensator (TCSC) and static VAR compensator. It was shown that falsified data at SCADA and PMU level alter system state variables, which in turn produce corrupt active and reactive power measurements at the TCSC terminal triggering incorrect control actions from the device. For attacks on static VAR, power injections at the terminal of the coupling transformer were altered to produce an incorrect firing angle. In both the cases, falsified data either drove the TCSC to operate very close to the resonance point or triggered the static VAR compensator to provide incorrect reactive power support. Such attacks have the potential to disturb the current operating conditions of the grid and physically damage expensive equipment.  

\section{Attacks on Substation Control}\label{sec:substation}

In this section, crafted attacks against substation measurement devices are discussed. These devices are responsible for sampling voltage, current and power flows from instrument transformers and relaying them to energy management system and other control and protection blocks. Attacks carried out at the device level are considered to be the most direct threats that obstruct real-time data flow, making operators blind to changes in the actual grid.

\subsection{Phasor Measurement Units}

Phasor measurement units, also known as synchrophasors, are widely deployed by utilities to improve real-time monitoring of voltage, current and frequency \cite{Phadke1993SynchronizedSystems,DeLaRee2010SynchronizedSystems}. These measurement units sample at the rate of 30/60/120 messages per second with a significantly high degree of accuracy compared to SCADA devices. PMUs at the substation level send their measurements to a local phasor data concentrator (PDC) where the data packets are time synchronized and aligned. Data from station level PDCs are concentrated at regional PDCs which further report to a centrally located data concentrator at the main control center. Any attack on the PMU-PDC architecture altering or blocking the flow of data packets can significantly impede various mission critical operations such as state estimation, real-time protection and control algorithms that rely on continuous streaming data. Various data attack models against PMU architecture have been discussed in the literature \cite{Zhang2013,Pal2016,Varmaziari2017Cyber-attackObserver,Yuan2011SecurityCounters}. Vulnerable points in the PMU-PDC architecture is shown in Fig. \ref{fig:pmuAttack}.

The first type of data injection attack is the manipulation of GPS time synchronization signal of the PMUs \cite{Zhang2013}. A global time stamp is assigned to each and every measurement sampled by the PMU. This enables operators to detect events, analyze faults and their locations in real-time. Modifying the time stamps associated with the PMU measurements is equivalent to altering phase angle of buses,
\begin{equation}
    \theta_{V_s}' =  \theta_{V_s} + \Delta \theta_S
\end{equation}
\begin{equation}
    \theta_{V_R}' =  \theta_{V_R} + \Delta \theta_R
\end{equation} 
where $\theta_{V_s}'$ and $\theta_{V_s}'$ are the modified sending and receiving end voltage phase angles after biases $\Delta \theta_S$ and $\Delta \theta_R$ are added to the original measurements. Such attacks can introduce locational errors in the event of fault location, which is given as,
\begin{equation}
    \Delta D = D - D(\Delta \theta)
\end{equation}
where $D$ is the original fault location index and $D(\Delta \theta)$ is the error introduced as a result of the attack. This significantly affects the system recovery process under contingencies.  Modification of the phase angle measurements also result in incorrect SE solutions. Additionally, when measurements at both ends of the line are tampered, an attacker can mask line outages. Incorrect topology results in falsified states affecting SCED, SCOPF and RTCA. 

Another type of vulnerability, called gray-hole attack, has been addressed by authors in \cite{Pal2016}. This kind of attack enable an adversary to drop PMU packet data in the PMU-PDC communication network. Three different metrics have been used to distinguish data packet drops due to network congestion from those dropped intentionally by an attacker: correlation metric, loss conditioned delay correlation and delay cumulative distribution function. The major impact of gray-hole attacks is the disruption of real-time data packet transmission which may result in topology errors and loss of measurements, ultimately leading to incorrect operator decisions at the control center. 

Man-in-the-Middle attack against PMUs have been studied by authors in \cite{Varmaziari2017Cyber-attackObserver}. Here, the operators are led to believe that messages received at the control center are from a trusted source. This kind of attack places the adversary between the measurement devices and the control center, subsequently allowing the attacker to modify PMU packet data. Another class of attack is code injection attack where malicious codes are inserted in the PMUs to add/delete/alter protection instructions \cite{Yuan2011SecurityCounters}.

\subsection{Threats Against IEC 61850}
IEC 61850 was introduced as the international standard for substation communication. It enables a simpler substation architecture and introduces interoperability between various commercially available protection and control devices \cite{Mackiewicz2006OverviewBenefits,Elgargouri2015IECSecurity}. This standard allows intelligent electronic devices (IEDs) to communicate with each other through local Ethernet using the Generic Object-Oriented Substation Event (GOOSE) messages \cite{Xu2013TowardsEnvironment,Sidhu2011ConfigurationGOOSE}. 

It was shown by authors in \cite{Rashid2014ANetwork,Kabir-Querrec2016ANetworks} that both the Ethernet and the GOOSE messaging protocols were vulnerable to data framing attacks. GOOSE messages exchanged between different control and monitoring equipment were demonstrated to be single point of failures vulnerable to attacks. These attacks mainly include data packet modification where the attackers can alter packet data contents over the LAN. Intentionally modified GOOSE messages, when received by control blocks, can disconnect circuit breakers, damage expensive field equipment, de-energize substations and cause multiple line outages. 

\subsection{Switching Attacks}

In \cite{liu2014coordinated}, authors have explored coordinated multi-switch attacks. These attacks remotely corrupt circuit breaker control signals through vulnerabilities in the communication networks. A variable structure system theory approach was used to study the behavior of power systems with switching dynamics. Attacks force circuit breakers to open or close for a short-duration that lead to transient destabilization of synchronous generators. This triggers protection systems to trip attacked generators. Three types of attacks were considered: single switching, concurrent switching and progressive switching attacks. The impacts of such attacks included loss of generation synchronization and subsequent tripping, loss of loads and uncontrolled cascades. A real world example of such an attack demonstration can be found here \cite{salmon2009mitigating}. 

\section{Energy Alteration Attacks}\label{sec:AMI_markets} 
In this section, attacks altering energy consumption and load forecast data are considered. These attacks forge meter measurements to alter demand information across a specific geographical area. Such attacks lead to incorrect operation of energy consumption scheduling blocks, increase in electricity prices for customers, incorrect micro-grid operations and economic profits for attackers from electricity markets.    

\subsection{Advanced Metering Infrastructures}
In smart grids, advanced metering infrastructures enable bi-directional communication between customers and utilities. AMI provides time-based information on electricity consumption, outages and electricity prices \cite{Selvam2012AdvancedApplications,Arian2011AdvancedArchitecture}.

Authors in \cite{Li2017HMM-BasedInfrastructure,Liu2015AInfrastructure} investigate false data attacks on smart meters, which form the backbone of advanced metering infrastructure. These attacks include Distributed Denial of Service (DDoS) attack that flood the meter communication network rendering critical meters unavailable. On the other hand, manipulation of smart meter data to claim higher energy export in an area can be exploited for economic gains \cite{Khanna2016DataProfit}. First, an attack region $N_{pri}$ involving buses that are not directly or indirectly connected to generators are identified. Once the attacking region is identified, the attackers maximize power injection at $i^{th}$ bus by solving the following optimization problem,
\begin{equation}
\begin{aligned}
& \underset{P_D^{N_{pri}},P_{Gi}}{\text{Max}}
& & \{ \sum_{l=1,l\neq i}^{N_{pri}} -B_{il}(\delta_l - \delta_i) \}\\
\end{aligned}
\end{equation} 
where $P_D$, $P_{Gi}$ and $\delta_i$ are the demand, real power generation and load angle at bus $i$ and $B$ is the bus susceptance matrix in the primary attack region. The optimization problem is solved while keeping altered load measurements within tolerable limits and ensuring that measurements in non-attacking region are not modified.

It can be perceived that attacks against AMI resemble load redistribution attacks. Falsifying power consumption and pricing data affect residential load controls, which further influence upper level control blocks such as SCOPF or SCED, as described in Section \ref{sec:steadyState}-C and \ref{sec:steadyState}-D. Moreover, when attackers gain access to ``remote disconnect” feature of the smart meters, they can disconnect a large group of customers causing widespread load shedding \cite{Grochocki2012AMIRecommendations}.

\subsection{Residential Load Control}

The energy consumption scheduling (ECS) and the residential load control systems are additional technologies that pave the way for bi-directional communication between consumers and distribution utilities \cite{Ullah2013ACommunications,Mohsenian-Rad2010AutonomousGrid}. The ECS system gathers customer consumption information and time-based electricity price to schedule energy for each household at each time \cite{Mohsenian-Rad2010OptimalEnvironments,Cohen1987AnControl}. This allows utilities to reliably supply electricity to all customers while minimizing cost of generated energy. 

False data attack on the ECS and the residential load control system lead to load alteration in the distribution grid, and hence change the output of the energy scheduling system. The problem of electricity price message fabrication has been studied by authors in \cite{Mishra2015RateGrid}. Two different indices have been studied to assess the impact of such an attack.

The first index calculates the maximum number of lines/nodes that can fail following a successful attack.  With $y_{ij}$ = 1 in case of line disconnection, $z_{i}$ = 1 if $i^{th}$ demand node is altered, $D$ denoting the set demand nodes, $(1+k_i)B$ as billing constraints, $r_i$ as the electricity billing rate and $(r_i-z_i.\rho_i)$ as the reduction in user demand at node $i$, the optimization problem was formulated as, \begin{equation}
\begin{aligned}\label{loadDistribute3}
& {\text{Max}}
& & \sum_{e(i,j) \in E} y_{ij}\\
&  \text{s.t.}& &  (1+k_i)B \geq D_i.(r_i-z_i.\rho_i)\\
& & & \sum_{i \in D} c_i(z_i) \leq A\\
& & & y_{ij} \le \frac{f_{ij}-u_{ij}}{u_{ij}}\\
\end{aligned}
\end{equation} The first constraint represents changes in demand values in billing calculation, the second constraint restricts the cost of attacking resources within attacker budget $A$ and the third constraint represents line disconnection. The optimization problem is subjected to additional power flow constraints. In the second index, the authors extended their work to study the failures of centrally located nodes. Attacks are launched with minimum cost burden on the attacker side while maximizing the potential of cascading failures.

In both the indices, the impact of altering load measurements has been quantified in terms of network stress, maximum available attack resources and network resilience to attacks. Such attacks at the consumer level can reflect adversely on the bulk transmission system. Single line failures may trigger additional line outages, which can eventually lead to a large portion of grid without power.  If the system is not resilient enough, such attacks might lead to cascading failures of critical edges and nodes.  

\subsection{Distributed Energy Routing}

Given multiple demand and supply node information as well as transmission line capacities, distributed energy routing algorithm finds the most optimal routes to transfer power from supply to demand nodes by minimizing the transmission costs \cite{Lin2013OnGrid}. This optimization problem for DER can be formulated as \cite{Lin2013OnGrid},
\begin{equation}
\begin{aligned}\label{distribute}
& {\text{min}}
& & cost=\frac{1}{2}.\sum_{L_{ij}\in L}(|E_{ij}.Cost_{ij}|)\\
& \text{s.t.} & &  \forall v \in N_P, \sum_{i \in N_v} E_{vi}\le P_v, \\
& & &  \forall u \in N_D, \sum_{j \in N_u} E_{uj}\le -D_u, \\
& & &  \forall L_{ij}\in L, E_{ij}=-E_{ji}  \\
\end{aligned}
\end{equation} where $i$ and $j$ are node numbers, $E_{ij}$ is the energy transmitted at $Cost_{ij}$ through transmission line $L_{ij} \in L$, where $L$ is the set of all lines, $N_P$ and $N_D$ are set of supply and demand nodes, $N_v$ and $N_u$ are the neighboring nodes of supply node $v$ and demand node $u$, $D_u$ is the node $u$ energy request and $P_v$ is the energy node $v$ can provide. 

Authors in \cite{Lin2012OnGrid} discuss data attacks against the distributed energy routing in smart grids. This kind of attack aims to manipulate demand and supply messages at different nodes. Three different scenarios have been considered to understand the impact of such attacks in terms of supplied energy loss, transmission cost and number of node outages. 

Under false energy data attack, the attacker falsifies node demand or supply data. False energy data attacks can be further classified into two types. In energy request deceiving attack, an attacker sends out large quantity of falsified energy request to different supply nodes. As the total energy supply is limited, compromised nodes demanding more energy will lead to other demand nodes not receiving their requested energy, and hence result in outages. Moreover, since the distributed energy routing algorithm is forced to re-route energy, it leads to increase in electricity transmission cost. On the other hand, in energy supply deceiving attack, an attacker claims to provide additional or less energy than under normal circumstances. This result in  increase in transmission costs and broken links, resulting in subsequent node outages as less nodes now receive requested energy.

Under false link-state attack, the attacker falsifies the transmission line state information. When out of service transmission lines are claimed to be valid, the energy distribution algorithm may assign power flow through these lines if they lie in the shortest path between demand and supply nodes. As the line is physically not able to carry energy, this would result in nodes outages and lead to transmission disruption. When valid lines are claimed to be invalid, the algorithm redirects energy through other non-optimal lines, increasing transmission cost. Claiming all lines to a node to be invalid results in further node outages.

\subsection{Micro-grids}\label{sec:microgrids} 

Micro-grids are small self-sustainable power grids local to communities or cities, which can disconnect from the main grid when required and operate autonomously. These micro-grids form self sufficient demand-supply systems often with large penetration of distributed renewable energy sources \cite{LasseterMicroGrids,Hatziargyriou2007Microgrids}. As micro-grids can operate independently, they strengthen grid resiliency at time of major contingencies. 

Authors in  \cite{Zhang2015OnGrid} investigate three different types of false data injection attacks affecting micro-grid partition. In such attacks, the attacker requests for forged energy than what is actually needed. The attacker can claim less energy than what a node can provide, or claim more energy than required, or a combination of both. When false energy is claimed, the process of dividing the grid into smaller micro-grids with required generation and load capacity fail. This imbalance results in the disruption of the dynamic partition process and results in node outages. 

Besides energy falsification, other data attack scenarios against micro-grids have been investigated in the literature. Authors in \cite{Chlela2016Real-timeAttacks} analyzed the total load lost due to frequency dips and system instability as a result of FDIA attacks. In \cite{Chlela2018FallbackCyber-Attacks}, authors have investigated scenarios where malicious firmware affect IEDs through a distributed denial-of-service attack. Such DDoS attacks alter real time active power levels of energy storage units causing frequency dips. Other attacks against micro grids involve disrupting the process of distributed load sharing control leading to instability of the system \cite{Zhang2019DistributedMicrogrid}. 
% Other attacks include attacks against substation voltage controller leading to mis-operation of micro-grids (hao)\textbf{[ref]}. 

\subsection{Electricity Markets}

Electricity market is another area in power system that is affected by energy forging attacks. Here the objective of the attacker is to make economic profit from electricity markets by altering Locational Marginal Prices (LMP). This is done by buying virtual power at lower prices and selling it at higher prices \cite{Xie2010FalseMarkets,Jia2012ImpactsOperations}. Reference \cite{Deng2017} provides comprehensive review of attacks on electricity markets.

The general structure of the electric market can be explained briefly as follows \cite{Xie2010FalseMarkets}: there are two different markets, the day-ahead (also called the ex-ante market) and the real-time market (ex-post market). To perform a successful attack, an attacker first buys and sells virtual power $P$ in the ex-ante market at different locations $j_1$ and $j_2$ with prices $\lambda^{DA}_{j1}$ and $\lambda^{DA}_{j2}$. Next, false data is injected to manipulate prices in the real-time market and the attacker sells and buys the virtual power $P$ at $j1$ and $j2$ with prices $\lambda_{j1}$ and $\lambda_{j2}$. The financial gains by the attacker can be formulated as, 
\begin{equation}
Profit=(\lambda_{j1}-\lambda_{j2}+\lambda^{DA}_{j2}-\lambda^{DA}_{j1})P  
\end{equation} Different from Locational Marginal Prices, authors in \cite{Lin2016TowardsMarkets} have discussed false data attacks on multistep electricity price considering forging power demand by compromising smart meters ultimately causing customers to pay more for electricity.

\section{Discussion}

This paper presented a taxonomy of nineteen different data attacks and discussed theoretical approaches of attack formulation. In this section, the general structure of data attack models and numerical software used to solve such models are reviewed. Emphasis is given on bi-level optimization that is commonly used to model attack scenarios. Further, discussions on current global cyber security practise and future research directions are also presented. 

\subsection{Mathematical Structures of Cyber Attacks in Power System}

Data attacks against state estimation are launched by exploiting the null-space of the system Jacobian matrix while compromising a minimum number of meters. To tackle the combinatorial problem of meter selection, different heuristics were considered. In \cite{QingyuYang2014OnCountermeasures}, a two level heuristic FDIA model was proposed  where a large network was first divided into smaller area. Next, brute force search was used to obtain reasonable sub-solutions for each area, which was then combined to obtain a global attack vector. In \cite{esmalifalak2011stealth}, linear independent component analysis techniques were used to infer topology and states. For FDIA with partial information, authors in \cite{rahman2012false} proposed a stochastic convex program that was solved using scenario generation methods. FDIA against automatic voltage controller was formulated using Partial Observable Markov Decision Process \cite{Chen2018EvaluationControl}.

In optimal attack vector designs for load redistribution attacks, topology falsification and coordinated cyber attacks, a similar mathematical construction was observed: bi-level optimization \cite{Yuan2011,xiang2015game,xiang2015power,Shelar2017CompromisingOperations,Liu2016MaskingAttacks,Khanna2017Bi-levelFlow}. Bi-level is a class of optimization problem where one optimization problem is nested inside another \cite{sinha2017review}. The general structure of bi-level program can be expressed as follows,

\begin{equation}
\begin{aligned}\label{distribute}
& {\text{minimize}_{x,y}}
& & f(x,y)\\
& \text{subject to} & &  h(x,y) \leq 0 \\
& & &  y = \text{arg min } g(x,y) \\
& & &   \hspace{1.2cm} \text{subject to } k(x,y) \leq 0\\
\end{aligned}
\end{equation}
\noindent 
where $x, y$ are upper and lower level variables respectively, $f(x,y)$ is the upper level objective function, $ g(x,y)$ is the lower level objective function, $h$ is the set of upper level constraints and $k$ is the set of lower level constraints. The lower level objective and its constraints together serve as constraints to the upper level problem. In a bi-level problem, feasible solutions include those that are optimal to lower level and satisfy all upper level constraints. The bi-level problem can be reformulated by replacing the lower level problem with the equivalent Karush-Kuhn-Tucker (KKT) conditions as,

\begin{equation}
\begin{aligned}\label{distribute}
& {\text{minimize}_{x,y, \lambda}}
& & f(x,y)\\
& \text{subject to} & &  h(x,y) \leq 0 \\
& & & k(x,y) \leq 0 \\
& & & \lambda_i \leq 0 & i=1,....,n \\
& & & \lambda_i k_i(x,y) = 0 & i=1,....,n \\
& & &  \triangledown_y \mathscr{L}(x,y,\lambda) = 0 \\
\end{aligned}
\end{equation}

\noindent where  $\mathscr{L}(x,y,\lambda) = g(x,y) + \sum_{i=1}^{n}\lambda_ik_i(x,y)$ is the lower level equivalent Lagrangian.

In \cite{Yuan2011}, load redistribution attack was formulated as a bi-level optimization problem. Here, the aim of the attacker was to inject false data to maximize generation and load-shedding costs while the operator aimed to reduce operation costs. Similarly, load redistribution attacks considering resource limitation \cite{xiang2015game,xiang2015power} and attacks against the economic dispatch \cite{Shelar2017CompromisingOperations} were formulated as bi-level problems. In all the above models, KKT conditions were used to transform the bi-level problem into a single level problem. Masking of transmission line outages using FDIA \cite{Liu2016MaskingAttacks} was also formulated as a bi-level problem. This was further reduced to a single level non-convex problem and was re-written as two linear programs. To solve the bi-level program for FDIA on security constrained optimal power flow, meta-heuristic techniques were employed for outer level optimization problem and quadrature programming was used to solve the inner level problem \cite{Khanna2017Bi-levelFlow}. Cooridnated cyber attack was formulated as mixed-integer nonlinear program and further reformulated as a mixed-integer linear program. Due to inherent non-convexity of lower level problem, a two-stage sequential approach was used instead of KKT based methods. In all the above models, commonly used optimization tools included CPLEX, Gurobi and GAMS.

All the above mathematical formulations demonstrate how attack vectors can be constructed from a theoretical perspective. Such attacks, involving alteration of measurements and status signals, translate to intrusion of secured networked systems in real life. To prevent unauthorized access, various 
cyber security policies for energy systems and 
critical infrastructures have been developed globally. The next sections highlights major global practices and policies.

\subsection{Global Cyber Security Practise}

The United Nations (UN), in its report on Protection of Critical Infrastructures (CIP), has laid down guidelines for best practices to combat cyber-security related issues \cite{unCTC}. Some of the broader guidelines of the CIP include (a) determining governmental agency responsible for overall implementation of CIP strategies, (b) risk assessment, identification and mitigation, (c) crisis management and developing appropriate responses to cyber threats, (d) incidence management and contingency plans for emergency events, (e) identification of threats at national, sector and company levels, (f) monitoring of industrial control system and intrusion detection, (g) technical security audits and (i) training and cyber drills. The UN Cybersecurity Programme offers technical assistance, capacity building and resources to nations worldwide for combatting cyber warfare against critical infrastructures. 

The National Cybersecurity Strategies (NSC), jointly developed by the International  Telecommunication Union, the World Bank, Commonwealth Secretariat, Commonwealth Telecommunications Organisation and NATO Cooperative Cyber Defence, houses a large number of existing cyber-security machinery and models \cite{itu}. One of the major aim of NSC is to create a common reference that would serve as a guide to countries in matters of cyber-security best practices and strategy development.

The North American Electric Reliability Corporation (NERC) has developed Critical Infrastructure Protection (CIP), 11 of which are currently subjected to enforcement \cite{Huff-Arkansas2011CIPStatus}. These CIP include (1)  categorization of security threats and identification of adverse impacts at different voltage levels of the bulk electric system, (2) security management controls, (3) personal and training, (4) electronic security perimeters, (5) physical security of cyber systems, (6) system security management, (7) incident reporting and response planning, (8) recovery plans, (9) configuration  change management and vulnerability assessments, (10) information protection and (11) physical security.

The China’s Cybersecurity Law \cite{protiviti} establishes the following major best practices which include: monitoring of network activity and storage of operation log, assignment of network security personnel, development of incident response plan, disaster recovery backup of critical data, remediation and incident response plan.

To address cyber security challenges in the Indian power sector, thirty four mandatory requirements (MR) have been proposed \cite{pillai2017indian}. The MR requires the appointment of a chief information security officer who is responsible for overall establishment of cyber security policies. Other MR include set up of information security department, formulating a framework for critical
information infrastructure, coordination with National Critical Information Infrastructure Protection Centre, dependency identification, critical asset inventory control, creation of electronic security perimeter, incident response management and security of redundant systems.

The Japanese Basic Act on Cybersecurity and its current amendment \cite{bac,lewis2015us} focuses on setting up of Cybersecurity Strategic Headquarters, monitoring of hostile activities against critical network infrastructures, cyber-security exercise and training, promotion of research at national universities, private sectors and research institutes, development of human resources and fostering international cooperation to successfully combat cyber crimes.

The Singapore Cybersecurity Act \cite{ter2018singapore} gives the Cybersecurity agency authority to secure government networks, establish methodology towards cyber-risk management, discover threats, develop incident response and tackle inter-dependencies between entities in case of an attack. The National Cybercrime Action Plan additionally aims to promote reliable data ecosystem, invest in cyber research and develop cyber security workforce.

The Australian Energy Market Operator (AEMO) commissioned Cyber Security Industry Working Group to develop the Australian Energy Sector Cyber Security Framework (AESCSF) \cite{AEMO}. The group consists of representatives from AEMO, industry and government and is chiefly responsible for developing criticality assessment tool and framework publication. An upcoming National Energy Cyber Security Readiness and Resilience Program aims to build a best practise framework for preventing, recognizing, responding and recovering from cyber attacks.

The African Union Commission (AUC) and Symantec \cite{africa} together put together cyber security guidelines that emphasis on stringent protection for single-point failures, firewall updates, monitoring network for intrusion detection, protecting private keys, implementing removable device policy, ensuring regular backups and proper incident response.

The European Electricity Subsector Cyber Security Capability Maturity Model \cite{ukparliament} identifies areas of smart grid where cyber security capability needs to be improved. Some of them include appointment of a central authority, incident reporting and information sharing on cyber attack patterns, creation of a certification board and setting up security standards.

\subsection{Future Research Directions}

With a surge in the number of cyber threats everyday, governments across nations are expanding their budget to secure their critical infrastructures. There is a rapid growth in the  current market for cyber security infrastructure. Both public and private enterprises are investing massive resources to secure smart grid data and network, restrict access control, identify, prevent zero day attacks and enable smarter intrusion detection. While a large number of studies have already demonstrated the potential of attacks originating in the cyber space and their damaging impact, the future of cyber security research lies in impact mitigation.
This becomes crucial as attacks get more sophisticated and harder to detect with existing technologies. Thus, more emphasis should be given on mitigation plans and how to isolate affected areas from the power network without impeding reliability. In case of uncertainties under attacks, robust defense strategies should be formulated and policies framed that aim to minimize the overall financial and social impact of cyber attack on the electric grid.

\section{Conclusion}

This paper presented a taxonomy of data attacks against four broad categories of power system operations and control blocks. Attacks against steady state control blocks affect static and dynamic state estimation, optimal power flow and security constrained economic dispatch. Attacks against transient and auxiliary control blocks threaten the integrity of automatic generation control, rotor angle stability controls, volt/VAR operations and FACTS devices. Substation control architectures fall prey to data attacks when PMU/SCADA architecture, IEC 61850 protocols and circuit breakers are targeted. Data alteration attacks also target smart meters, load control blocks, micro-grids and electricity markets. For each category of attack, mathematical models are discussed. Emphasis is given on bi-level optimization framework which is commonly used to construct optimal attacks. Additionally, a chronology of benchmark research in areas of power system cyber security attacks and global cyber security policies are presented. All such data falsification attacks can result in incorrect operator decisions, uneconomic operations, physical line outages and voltage collapse, thereby increasing the potential of cascading failures. Overall, our taxonomy of data attacks provides a structured approach to help researchers, operators and policy makers understand the imminent cyber threats faced by modern power systems. 

% \textbf{Provide IEEE definitions for each elements in the beginning of the paragraph}
 
\bibliographystyle{IEEEtran}
\bibliography{References}

\begin{thebibliography}{100}
\providecommand{\url}[1]{#1}
\csname url@samestyle\endcsname
\providecommand{\newblock}{\relax}
\providecommand{\bibinfo}[2]{#2}
\providecommand{\BIBentrySTDinterwordspacing}{\spaceskip=0pt\relax}
\providecommand{\BIBentryALTinterwordstretchfactor}{4}
\providecommand{\BIBentryALTinterwordspacing}{\spaceskip=\fontdimen2\font plus
\BIBentryALTinterwordstretchfactor\fontdimen3\font minus
  \fontdimen4\font\relax}
\providecommand{\BIBforeignlanguage}[2]{{%
\expandafter\ifx\csname l@#1\endcsname\relax
\typeout{** WARNING: IEEEtran.bst: No hyphenation pattern has been}%
\typeout{** loaded for the language `#1'. Using the pattern for}%
\typeout{** the default language instead.}%
\else
\language=\csname l@#1\endcsname
\fi
#2}}
\providecommand{\BIBdecl}{\relax}
\BIBdecl

\bibitem{CherepanovAnton}
\BIBentryALTinterwordspacing
A.~Cherepanov and R.~Lipovsky., ``{Industroyer: Biggest threat to industrial
  control systems since Stuxnet},'' 2017. [Online]. Available:
  \url{WeLiveSecurity, ESET 12 (2017).}
\BIBentrySTDinterwordspacing

\bibitem{Liang2017TheAttacks}
G.~Liang, S.~R. Weller, J.~Zhao, F.~Luo, and Z.~Y. Dong, ``{The 2015 Ukraine
  Blackout: Implications for False Data Injection Attacks},'' \emph{IEEE
  Transactions on Power Systems}, vol.~32, no.~4, pp. 3317--3318, 7 2017.

\bibitem{2014Dragonfly:Response}
``{Dragonfly: Cyberespionage Attacks Against Energy Suppliers Symantec Security
  Response},'' Tech. Rep., 2014.

\bibitem{Chen2017AutomatedRansomware}
Q.~Chen and R.~A. Bridges, ``{Automated Behavioral Analysis of Malware: A Case
  Study of WannaCry Ransomware},'' in \emph{2017 16th IEEE International
  Conference on Machine Learning and Applications (ICMLA)}.\hskip 1em plus
  0.5em minus 0.4em\relax IEEE, 12 2017, pp. 454--460.

\bibitem{Hsiao2018TheRansomware}
S.-C. Hsiao and D.-Y. Kao, ``{The static analysis of WannaCry ransomware},'' in
  \emph{2018 20th International Conference on Advanced Communication Technology
  (ICACT)}.\hskip 1em plus 0.5em minus 0.4em\relax IEEE, 2 2018, pp. 153--158.

\bibitem{Langner2011Stuxnet:Weapon}
R.~Langner, ``{Stuxnet: Dissecting a Cyberwarfare Weapon},'' \emph{IEEE
  Security {\&} Privacy Magazine}, vol.~9, no.~3, pp. 49--51, 5 2011.

\bibitem{Karnouskos2011StuxnetSecurity}
S.~Karnouskos, ``{Stuxnet worm impact on industrial cyber-physical system
  security},'' in \emph{IECON 2011 - 37th Annual Conference of the IEEE
  Industrial Electronics Society}.\hskip 1em plus 0.5em minus 0.4em\relax IEEE,
  11 2011, pp. 4490--4494.

\bibitem{Chen2011LessonsStuxnet}
T.~M. Chen and S.~Abu-Nimeh, ``{Lessons from Stuxnet},'' \emph{Computer},
  vol.~44, no.~4, pp. 91--93, 4 2011.

\bibitem{ThreatSTOP2016BlackEnergyReport}
{ThreatSTOP}, ``{BlackEnergy Security report},'' Tech. Rep., 2016.

\bibitem{dragosTrisis}
Dragos, ``Trisis malware analysis of safety system targeted malware,'' 2019.

\bibitem{ComputerMcAfee}
``{Computer Virus Attacks, Information, News, Security, Detection and Removal |
  McAfee}.''

\bibitem{Stamp2009SANDIAFY08}
J.~E. Stamp, R.~A. Laviolette, L.~R. Phillips, and B.~T. Richardson, ``{Sandia
  Final Report: Impacts Analysis for Cyber Attack on Electric Power Systems
  (National SCADA Test Bed FY08)},'' Tech. Rep., 2009.

\bibitem{TheSmartGridInteroperabilityPanel2010IntroductionSecurity}
{The Smart Grid Interoperability Panel} and {Cyber Security Working Group},
  ``{Introduction to NISTIR 7628 Guidelines for Smart Grid Cyber Security},''
  Tech. Rep., 2010.

\bibitem{Huff-Arkansas2011CIPStatus}
P.~Huff~Arkansas and D.~Revill~Georgia, ``{CIP Standards Version 5 Status and
  Requirements },'' Tech. Rep., 2011.

\bibitem{NERCCyberReport}
{NERC}, ``{Cyber Attack Task Force Final Report},'' Tech. Rep.

\bibitem{2010High-ImpactSystem}
``{High-Impact, Low-Frequency Event Risk to the North American Bulk Power
  System},'' Tech. Rep., 2010.

\bibitem{Campbell2015CybersecuritySystem}
R.~J. Campbell, ``{Cybersecurity Issues for the Bulk Power System},'' Tech.
  Rep., 2015.

\bibitem{IdahoNationalLaboratory2017}
{Idaho National Laboratory}, ``{Cyber Threat and Vulnerability Analysis of the
  U.S. Electric Sector},'' Idaho National Laboratory, Tech. Rep., 2017.

\bibitem{Schlichting2018AssessmentOSD}
A.~D. Schlichting, ``{Assessment of Operational Energy System Cybersecurity
  Vulnerabilities Sponsor: OSD},'' Tech. Rep., 2018.

\bibitem{Yan2013AChallenges}
Y.~Yan, Y.~Qian, H.~Sharif, and D.~Tipper, ``{A Survey on Smart Grid
  Communication Infrastructures: Motivations, Requirements and Challenges},''
  \emph{IEEE Communications Surveys {\&} Tutorials}, vol.~15, no.~1, pp. 5--20,
  21 2013.

\bibitem{Ericsson2010CyberInfrastructure}
G.~N. Ericsson, ``{Cyber Security and Power System Communication—Essential
  Parts of a Smart Grid Infrastructure},'' \emph{IEEE Transactions on Power
  Delivery}, vol.~25, no.~3, pp. 1501--1507, 7 2010.

\bibitem{Yan2012ACommunications}
Y.~Yan, Y.~Qian, H.~Sharif, and D.~Tipper, ``{A Survey on Cyber Security for
  Smart Grid Communications},'' \emph{IEEE Communications Surveys {\&}
  Tutorials}, vol.~14, no.~4, pp. 998--1010, 24 2012.

\bibitem{YilinMo2012CyberPhysicalInfrastructure}
{Yilin Mo}, T.~H.-J. Kim, K.~Brancik, D.~Dickinson, {Heejo Lee}, A.~Perrig, and
  B.~Sinopoli, ``{Cyber–Physical Security of a Smart Grid Infrastructure},''
  \emph{Proceedings of the IEEE}, vol. 100, no.~1, pp. 195--209, 1 2012.

\bibitem{Mrabet2018Cyber-securityChallenges}
Z.~E. Mrabet, N.~Kaabouch, H.~E. Ghazi, and H.~E. Ghazi, ``{Cyber-security in
  smart grid: Survey and challenges},'' \emph{Computers {\&} Electrical
  Engineering}, vol.~67, pp. 469--482, 4 2018.

\bibitem{Li2012SecuringChallenges}
X.~Li, X.~Liang, R.~Lu, X.~Shen, X.~Lin, and H.~Zhu, ``{Securing smart grid:
  cyber attacks, countermeasures, and challenges},'' \emph{IEEE Communications
  Magazine}, vol.~50, no.~8, pp. 38--45, 8 2012.

\bibitem{Jokar2016AGrids}
P.~Jokar, N.~Arianpoo, and V.~C.~M. Leung, ``{A survey on security issues in
  smart grids},'' \emph{Security and Communication Networks}, vol.~9, no.~3,
  pp. 262--273, 2 2016.

\bibitem{Liu2012CyberGrids}
J.~Liu, Y.~Xiao, S.~Li, W.~Liang, and C.~L.~P. Chen, ``{Cyber Security and
  Privacy Issues in Smart Grids},'' \emph{IEEE Communications Surveys {\&}
  Tutorials}, vol.~14, no.~4, pp. 981--997, 24 2012.

\bibitem{Sridhar2012CyberPhysicalGrid}
S.~Sridhar, A.~Hahn, and M.~Govindarasu, ``{Cyber–Physical System Security
  for the Electric Power Grid},'' \emph{Proceedings of the IEEE}, vol. 100,
  no.~1, pp. 210--224, 1 2012.

\bibitem{Sun2018CyberState-of-the-art}
C.-C. Sun, A.~Hahn, and C.-C. Liu, ``{Cyber security of a power grid:
  State-of-the-art},'' \emph{International Journal of Electrical Power {\&}
  Energy Systems}, vol.~99, pp. 45--56, 7 2018.

\bibitem{Wang2013AGrid}
D.~Wang, X.~Guan, T.~Liu, Y.~Gu, Y.~Sun, and Y.~Liu, ``{A survey on bad data
  injection attack in smart grid},'' in \emph{2013 IEEE PES Asia-Pacific Power
  and Energy Engineering Conference (APPEEC)}.\hskip 1em plus 0.5em minus
  0.4em\relax IEEE, 12 2013, pp. 1--6.

\bibitem{Pour2017ASystems}
M.~M. Pour, A.~Anzalchi, and A.~Sarwat, ``{A review on cyber security issues
  and mitigation methods in smart grid systems},'' in \emph{SoutheastCon
  2017}.\hskip 1em plus 0.5em minus 0.4em\relax IEEE, 3 2017, pp. 1--4.

\bibitem{Rasmussen2017AAssessment}
T.~B. Rasmussen, G.~Yang, A.~H. Nielsen, and Z.~Dong, ``{A review of
  cyber-physical energy system security assessment},'' in \emph{2017 IEEE
  Manchester PowerTech}.\hskip 1em plus 0.5em minus 0.4em\relax IEEE, 6 2017,
  pp. 1--6.

\bibitem{Line2011CyberGrids}
M.~B. Line, I.~A. Tondel, and M.~G. Jaatun, ``{Cyber security challenges in
  Smart Grids},'' in \emph{2011 2nd IEEE PES International Conference and
  Exhibition on Innovative Smart Grid Technologies}.\hskip 1em plus 0.5em minus
  0.4em\relax IEEE, 12 2011, pp. 1--8.

\bibitem{Kotut2016SurveyGrids}
L.~Kotut and L.~A. Wahsheh, ``{Survey of Cyber Security Challenges and
  Solutions in Smart Grids},'' in \emph{2016 Cybersecurity Symposium
  (CYBERSEC)}.\hskip 1em plus 0.5em minus 0.4em\relax IEEE, 4 2016, pp. 32--37.

\bibitem{Chatterjee2017ReviewOperations}
K.~Chatterjee, V.~Padmini, and S.~A. Khaparde, ``{Review of cyber attacks on
  power system operations},'' in \emph{2017 IEEE Region 10 Symposium
  (TENSYMP)}.\hskip 1em plus 0.5em minus 0.4em\relax IEEE, 7 2017, pp. 1--6.

\bibitem{Deng2017}
R.~Deng, G.~Xiao, R.~Lu, H.~Liang, and A.~V. Vasilakos, ``{False Data Injection
  on State Estimation in Power Systems—Attacks, Impacts, and Defense: A
  Survey},'' \emph{IEEE Transactions on Industrial Informatics}, vol.~13,
  no.~2, pp. 411--423, 4 2017.

\bibitem{Liang2017}
G.~Liang, J.~Zhao, F.~Luo, S.~R. Weller, and Z.~Y. Dong, ``{A Review of False
  Data Injection Attacks Against Modern Power Systems},'' \emph{IEEE
  Transactions on Smart Grid}, vol.~8, no.~4, pp. 1630--1638, 7 2017.

\bibitem{Zhang2018CanSystems}
J.~Zhang, Z.~Chu, L.~Sankar, and O.~Kosut, ``{Can Attackers With Limited
  Information Exploit Historical Data to Mount Successful False Data Injection
  Attacks on Power Systems?}'' \emph{IEEE Transactions on Power Systems},
  vol.~33, no.~5, pp. 4775--4786, 9 2018.

\bibitem{Liu2009}
Y.~Liu, M.~K. Reiter, P.~Ning, and M.~K. Reiter, ``{False data injection
  attacks against state estimation in electric power grids},'' in
  \emph{Proceedings of the 16th ACM conference on Computer and communications
  security - CCS '09}, vol.~14, no.~1.\hskip 1em plus 0.5em minus 0.4em\relax
  New York, New York, USA: ACM Press, 5 2009, p.~21.

\bibitem{QingyuYang2014OnCountermeasures}
{Qingyu Yang}, {Jie Yang}, {Wei Yu}, {Dou An}, {Nan Zhang}, and {Wei Zhao},
  ``{On False Data-Injection Attacks against Power System State Estimation:
  Modeling and Countermeasures},'' \emph{IEEE Transactions on Parallel and
  Distributed Systems}, vol.~25, no.~3, pp. 717--729, 3 2014.

\bibitem{Hug2012}
G.~Hug and J.~A. Giampapa, ``{Vulnerability assessment of AC state estimation
  with respect to false data injection cyber-attacks},'' \emph{IEEE
  Transactions on Smart Grid}, vol.~3, no.~3, pp. 1362--1370, 9 2012.

\bibitem{Liu2017FalseInformation}
X.~Liu and Z.~Li, ``{False Data Attacks Against AC State Estimation With
  Incomplete Network Information},'' \emph{IEEE Transactions on Smart Grid},
  vol.~8, no.~5, pp. 2239--2248, 9 2017.

\bibitem{esmalifalak2011stealth}
M.~Esmalifalak, H.~Nguyen, R.~Zheng, and Z.~Han, ``Stealth false data injection
  using independent component analysis in smart grid,'' in \emph{Smart Grid
  Communications (SmartGridComm), 2011 IEEE International Conference on}.\hskip
  1em plus 0.5em minus 0.4em\relax IEEE, 2011, pp. 244--248.

\bibitem{rahman2012false}
M.~A. Rahman and H.~Mohsenian-Rad, ``False data injection attacks with
  incomplete information against smart power grids,'' in \emph{Global
  Communications Conference (GLOBECOM), 2012 IEEE}.\hskip 1em plus 0.5em minus
  0.4em\relax Citeseer, 2012, pp. 3153--3158.

\bibitem{Yu2015BlindGrid}
Z.-H. Yu and W.-L. Chin, ``{Blind False Data Injection Attack Using PCA
  Approximation Method in Smart Grid},'' \emph{IEEE Transactions on Smart
  Grid}, vol.~6, no.~3, pp. 1219--1226, 5 2015.

\bibitem{Basumallik2017}
S.~Basumallik, S.~Eftekharnejad, N.~Davis, and B.~B.~K. Johnson, ``{Impact of
  false data injection attacks on PMU-based state estimation}.''\hskip 1em plus
  0.5em minus 0.4em\relax IEEE, 9 2017, pp. 1--6.

\bibitem{Chen2017AGrid}
R.~Chen, D.~Du, and M.~Fei, ``{A Novel Data Injection Cyber-Attack Against
  Dynamic State Estimation in Smart Grid}.''\hskip 1em plus 0.5em minus
  0.4em\relax Springer, Singapore, 2017, pp. 607--615.

\bibitem{Khanna2017Bi-levelFlow}
K.~Khanna, B.~K. Panigrahi, and A.~Joshi, ``{Bi-level modelling of false data
  injection attacks on security constrained optimal power flow},'' \emph{IET
  Generation, Transmission {\&} Distribution}, vol.~11, no.~14, pp. 3586--3593,
  9 2017.

\bibitem{Shelar2017CompromisingOperations}
D.~Shelar, P.~Sun, S.~Amin, and S.~Zonouz, ``{Compromising Security of Economic
  Dispatch in Power System Operations},'' in \emph{2017 47th Annual IEEE/IFIP
  International Conference on Dependable Systems and Networks (DSN)}.\hskip 1em
  plus 0.5em minus 0.4em\relax IEEE, 6 2017, pp. 531--542.

\bibitem{Yuan2011}
Y.~Yuan, Z.~Li, and K.~Ren, ``{Modeling Load Redistribution Attacks in Power
  Systems},'' \emph{IEEE Transactions on Smart Grid}, vol.~2, no.~2, pp.
  382--390, 6 2011.

\bibitem{yuan2012quantitative}
{Yuan, Yanling}, {Li, Zuyi}, and {Ren, Kui}, ``Quantitative analysis of load
  redistribution attacks in power systems,'' \emph{IEEE Transactions on
  Parallel and Distributed Systems}, vol.~23, no.~9, pp. 1731--1738, 2012.

\bibitem{Liu2014LocalInformation}
X.~Liu and Z.~Li, ``{Local Load Redistribution Attacks in Power Systems With
  Incomplete Network Information},'' \emph{IEEE Transactions on Smart Grid},
  vol.~5, no.~4, pp. 1665--1676, 7 2014.

\bibitem{Liu2015ModelingInformation}
X.~Liu, Z.~Bao, D.~Lu, and Z.~Li, ``{Modeling of Local False Data Injection
  Attacks With Reduced Network Information},'' \emph{IEEE Transactions on Smart
  Grid}, vol.~6, no.~4, pp. 1686--1696, 7 2015.

\bibitem{li2018analyzing}
Z.~Li, M.~Shahidehpour, A.~Alabdulwahab, and A.~Abusorrah, ``Analyzing locally
  coordinated cyber-physical attacks for undetectable line outages,''
  \emph{IEEE Transactions on Smart Grid}, vol.~9, no.~1, pp. 35--47, 2018.

\bibitem{xiang2015power}
Y.~Xiang, Z.~Ding, and L.~Wang, ``Power system adequacy assessment with load
  redistribution attacks,'' in \emph{Innovative Smart Grid Technologies
  Conference (ISGT), 2015 IEEE Power \& Energy Society}.\hskip 1em plus 0.5em
  minus 0.4em\relax IEEE, 2015, pp. 1--5.

\bibitem{xiang2015game}
Y.~Xiang and L.~Wang, ``A game-theoretic approach to optimal defense strategy
  against load redistribution attack,'' in \emph{Power \& Energy Society
  General Meeting, 2015 IEEE}.\hskip 1em plus 0.5em minus 0.4em\relax IEEE,
  2015, pp. 1--5.

\bibitem{xiang2017framework}
Y.~Xiang, L.~Wang, and N.~Liu, ``A framework for modeling load redistribution
  attacks coordinating with switching attacks,'' in \emph{Power \& Energy
  Society General Meeting, 2017 IEEE}.\hskip 1em plus 0.5em minus 0.4em\relax
  IEEE, 2017, pp. 1--5.

\bibitem{xiang2015coordinated}
Y.~Xiang, L.~Wang, D.~Yu, and N.~Liu, ``Coordinated attacks against power
  grids: Load redistribution attack coordinating with generator and line
  attacks,'' in \emph{Power \& Energy Society General Meeting, 2015
  IEEE}.\hskip 1em plus 0.5em minus 0.4em\relax IEEE, 2015, pp. 1--5.

\bibitem{pinceti2018load}
A.~Pinceti, L.~Sankar, and O.~Kosut, ``Load redistribution attack detection
  using machine learning: A data-driven approach,'' in \emph{2018 IEEE Power \&
  Energy Society General Meeting (PESGM)}.\hskip 1em plus 0.5em minus
  0.4em\relax IEEE, 2018, pp. 1--5.

\bibitem{Liu2016MaskingAttacks}
X.~Liu, Z.~Li, X.~Liu, and Z.~Li, ``{Masking Transmission Line Outages via
  False Data Injection Attacks},'' \emph{IEEE Transactions on Information
  Forensics and Security}, vol.~11, no.~7, pp. 1592--1602, 7 2016.

\bibitem{Kim2013OnCountermeasures}
J.~Kim and L.~Tong, ``{On Topology Attack of a Smart Grid: Undetectable Attacks
  and Countermeasures},'' \emph{IEEE Journal on Selected Areas in
  Communications}, vol.~31, no.~7, pp. 1294--1305, 7 2013.

\bibitem{lin2013distributed}
J.~Lin, W.~Yu, D.~Griffith, X.~Yang, G.~Xu, and C.~Lu, ``On distributed energy
  routing protocols in the smart grid,'' in \emph{Software Engineering,
  Artificial Intelligence, Networking and Parallel/Distributed
  Computing}.\hskip 1em plus 0.5em minus 0.4em\relax Springer, 2013, pp.
  143--159.

\bibitem{Li2016BilevelSystems}
Z.~Li, M.~Shahidehpour, A.~Alabdulwahab, and A.~Abusorrah, ``{Bilevel Model for
  Analyzing Coordinated Cyber-Physical Attacks on Power Systems},'' \emph{IEEE
  Transactions on Smart Grid}, vol.~7, no.~5, pp. 2260--2272, 9 2016.

\bibitem{Deng2017CCPA:Grid}
R.~Deng, P.~Zhuang, and H.~Liang, ``{CCPA: Coordinated Cyber-Physical Attacks
  and Countermeasures in Smart Grid},'' \emph{IEEE Transactions on Smart Grid},
  vol.~8, no.~5, pp. 2420--2430, 9 2017.

\bibitem{Farraj2017OnControl}
A.~Farraj, E.~Hammad, and D.~Kundur, ``{On the Impact of Cyber Attacks on Data
  Integrity in Storage-Based Transient Stability Control},'' \emph{IEEE
  Transactions on Industrial Informatics}, vol.~13, no.~6, pp. 3322--3333, 12
  2017.

\bibitem{Sridhar2014Model-BasedControl}
S.~Sridhar and M.~Govindarasu, ``{Model-Based Attack Detection and Mitigation
  for Automatic Generation Control},'' \emph{IEEE Transactions on Smart Grid},
  vol.~5, no.~2, pp. 580--591, 3 2014.

\bibitem{Ashok2015ExperimentalTestbed}
A.~Ashok, {Pengyuan Wang}, M.~Brown, and M.~Govindarasu, ``{Experimental
  evaluation of cyber attacks on Automatic Generation Control using a CPS
  Security Testbed},'' in \emph{2015 IEEE Power {\&} Energy Society General
  Meeting}.\hskip 1em plus 0.5em minus 0.4em\relax IEEE, 7 2015, pp. 1--5.

\bibitem{Tan2017ModelingControl}
R.~Tan, H.~H. Nguyen, E.~Y.~S. Foo, D.~K.~Y. Yau, Z.~Kalbarczyk, R.~K. Iyer,
  and H.~B. Gooi, ``{Modeling and Mitigating Impact of False Data Injection
  Attacks on Automatic Generation Control},'' \emph{IEEE Transactions on
  Information Forensics and Security}, vol.~12, no.~7, pp. 1609--1624, 7 2017.

\bibitem{Chen2018EvaluationControl}
Y.~Chen, S.~Huang, F.~Liu, Z.~Wang, and X.~Sun, ``{Evaluation of Reinforcement
  Learning Based False Data Injection Attack to Automatic Voltage Control},''
  \emph{IEEE Transactions on Smart Grid}, pp. 1--1, 2018.

\bibitem{Teixeira2014SecurityCountermeasures}
A.~Teixeira, G.~Dan, H.~Sandberg, R.~Berthier, R.~B. Bobba, and A.~Valdes,
  ``{Security of smart distribution grids: Data integrity attacks on integrated
  volt/VAR control and countermeasures},'' in \emph{2014 American Control
  Conference}.\hskip 1em plus 0.5em minus 0.4em\relax IEEE, 6 2014, pp.
  4372--4378.

\bibitem{Rubio-Marroquin2018ImpactParameters}
G.~O. Rubio-Marroquin, C.~R. Fuerte-Esquivel, and E.~A. Zamora-Cardenas,
  ``{Impact of bad data injection attacks in the estimation of FACTS
  controllers parameters},'' in \emph{2018 IEEE Power {\&} Energy Society
  Innovative Smart Grid Technologies Conference (ISGT)}.\hskip 1em plus 0.5em
  minus 0.4em\relax IEEE, 2 2018, pp. 1--5.

\bibitem{Zhang2013}
Z.~Zhang, S.~Gong, A.~D. Dimitrovski, and H.~Li, ``{Time Synchronization Attack
  in Smart Grid: Impact and Analysis},'' \emph{IEEE Transactions on Smart
  Grid}, vol.~4, no.~1, pp. 87--98, 3 2013.

\bibitem{Pal2016}
S.~Pal, B.~Sikdar, and J.~Chow, ``{An Online Mechanism for Detection of
  Gray-Hole Attacks on PMU Data},'' \emph{IEEE Transactions on Smart Grid}, pp.
  1--1, 2018.

\bibitem{Varmaziari2017Cyber-attackObserver}
H.~Varmaziari and M.~Dehghani, ``{Cyber-attack detection system of large-scale
  power systems using decentralized unknown input observer},'' in \emph{2017
  Iranian Conference on Electrical Engineering (ICEE)}.\hskip 1em plus 0.5em
  minus 0.4em\relax IEEE, 5 2017, pp. 621--626.

\bibitem{Yuan2011SecurityCounters}
L.~Yuan, W.~Xing, H.~Chen, and B.~Zang, ``{Security Breaches as PMU Deviation:
  Detecting and Identifying Security Attacks Using Performance Counters},'' in
  \emph{ACM SIGOPS Asia-pacific Workshop on Systems}, 2011.

\bibitem{Rashid2014ANetwork}
M.~T.~A. Rashid, S.~Yussof, Y.~Yusoff, and R.~Ismail, ``{A review of security
  attacks on IEC61850 substation automation system network},'' in
  \emph{Proceedings of the 6th International Conference on Information
  Technology and Multimedia}.\hskip 1em plus 0.5em minus 0.4em\relax IEEE, 11
  2014, pp. 5--10.

\bibitem{Kabir-Querrec2016ANetworks}
M.~Kabir-Querrec, S.~Mocanu, J.-M. Thiriet, and E.~Savary, ``{A Test bed
  dedicated to the Study of Vulnerabilities in IEC 61850 Power Utility
  Automation Networks},'' in \emph{2016 IEEE 21st International Conference on
  Emerging Technologies and Factory Automation (ETFA)}.\hskip 1em plus 0.5em
  minus 0.4em\relax IEEE, 9 2016, pp. 1--4.

\bibitem{liu2014coordinated}
S.~Liu, B.~Chen, T.~Zourntos, D.~Kundur, and K.~Butler-Purry, ``A coordinated
  multi-switch attack for cascading failures in smart grid,'' \emph{IEEE
  Transactions on Smart Grid}, vol.~5, no.~3, pp. 1183--1195, 2014.

\bibitem{Li2017HMM-BasedInfrastructure}
B.~Li, R.~Lu, and G.~Xiao, ``{HMM-Based Fast Detection of False Data Injections
  in Advanced Metering Infrastructure},'' in \emph{GLOBECOM 2017 - 2017 IEEE
  Global Communications Conference}.\hskip 1em plus 0.5em minus 0.4em\relax
  IEEE, 12 2017, pp. 1--6.

\bibitem{Liu2015AInfrastructure}
X.~Liu, P.~Zhu, Y.~Zhang, and K.~Chen, ``{A Collaborative Intrusion Detection
  Mechanism Against False Data Injection Attack in Advanced Metering
  Infrastructure},'' \emph{IEEE Transactions on Smart Grid}, vol.~6, no.~5, pp.
  2435--2443, 9 2015.

\bibitem{Khanna2016DataProfit}
K.~Khanna, B.~K. Panigrahi, and A.~Joshi, ``{Data integrity attack in smart
  grid: optimised attack to gain momentary economic profit},'' \emph{IET
  Generation, Transmission {\&} Distribution}, vol.~10, no.~16, pp. 4032--4039,
  12 2016.

\bibitem{Grochocki2012AMIRecommendations}
D.~Grochocki, J.~H. Huh, R.~Berthier, R.~Bobba, W.~H. Sanders, A.~A. Cardenas,
  and J.~G. Jetcheva, ``{AMI threats, intrusion detection requirements and
  deployment recommendations},'' in \emph{2012 IEEE Third International
  Conference on Smart Grid Communications (SmartGridComm)}.\hskip 1em plus
  0.5em minus 0.4em\relax IEEE, 11 2012, pp. 395--400.

\bibitem{Mishra2015RateGrid}
S.~Mishra, X.~Li, A.~Kuhnle, M.~T. Thai, and J.~Seo, ``{Rate alteration attacks
  in smart grid},'' in \emph{2015 IEEE Conference on Computer Communications
  (INFOCOM)}.\hskip 1em plus 0.5em minus 0.4em\relax IEEE, 4 2015, pp.
  2353--2361.

\bibitem{Lin2012OnGrid}
\BIBentryALTinterwordspacing
J.~Lin, W.~Yu, X.~Yang, G.~Xu, and W.~Zhao, ``{On False Data Injection Attacks
  against Distributed Energy Routing in Smart Grid},'' in \emph{2012 IEEE/ACM
  Third International Conference on Cyber-Physical Systems}.\hskip 1em plus
  0.5em minus 0.4em\relax IEEE, 4 2012, pp. 183--192. [Online]. Available:
  \url{http://ieeexplore.ieee.org/document/6197400/}
\BIBentrySTDinterwordspacing

\bibitem{Zhang2015OnGrid}
X.~Zhang, X.~Yang, J.~Lin, and W.~Yu, ``{On false data injection attacks
  against the dynamic microgrid partition in the smart grid},'' in \emph{2015
  IEEE International Conference on Communications (ICC)}.\hskip 1em plus 0.5em
  minus 0.4em\relax IEEE, 6 2015, pp. 7222--7227.

\bibitem{Chlela2016Real-timeAttacks}
M.~Chlela, G.~Joos, M.~Kassouf, and Y.~Brissette, ``{Real-time testing platform
  for microgrid controllers against false data injection cybersecurity
  attacks},'' in \emph{2016 IEEE Power and Energy Society General Meeting
  (PESGM)}.\hskip 1em plus 0.5em minus 0.4em\relax IEEE, 7 2016, pp. 1--5.

\bibitem{Chlela2018FallbackCyber-Attacks}
M.~Chlela, D.~Mascarella, G.~Joos, and M.~Kassouf, ``{Fallback Control for
  Isochronous Energy Storage Systems in Autonomous Microgrids Under
  Denial-of-Service Cyber-Attacks},'' \emph{IEEE Transactions on Smart Grid},
  vol.~9, no.~5, pp. 4702--4711, 9 2018.

\bibitem{Xie2010FalseMarkets}
L.~Xie, Y.~Mo, and B.~Sinopoli, ``{False Data Injection Attacks in Electricity
  Markets},'' in \emph{2010 First IEEE International Conference on Smart Grid
  Communications}.\hskip 1em plus 0.5em minus 0.4em\relax IEEE, 10 2010, pp.
  226--231.

\bibitem{Lin2016TowardsMarkets}
J.~Lin, W.~Yu, and X.~Yang, ``{Towards Multistep Electricity Prices in Smart
  Grid Electricity Markets},'' \emph{IEEE Transactions on Parallel and
  Distributed Systems}, vol.~27, no.~1, pp. 286--302, 1 2016.

\bibitem{Karimipour2017OnGrids}
H.~Karimipour and V.~Dinavahi, ``{On false data injection attack against
  dynamic state estimation on smart power grids},'' in \emph{2017 IEEE
  International Conference on Smart Energy Grid Engineering (SEGE)}.\hskip 1em
  plus 0.5em minus 0.4em\relax IEEE, 8 2017, pp. 388--393.

\bibitem{Zhang2019DistributedMicrogrid}
H.~Zhang, W.~Meng, J.~Qi, X.~Wang, and W.~X. Zheng, ``{Distributed Load Sharing
  Under False Data Injection Attack in an Inverter-Based Microgrid},''
  \emph{IEEE Transactions on Industrial Electronics}, vol.~66, no.~2, pp.
  1543--1551, 2 2019.

\bibitem{Schweppe1970PowerModel}
F.~Schweppe and J.~Wildes, ``{Power System Static-State Estimation, Part I:
  Exact Model},'' \emph{IEEE Transactions on Power Apparatus and Systems}, vol.
  PAS-89, no.~1, pp. 120--125, 1 1970.

\bibitem{Jaleeli1992UnderstandingControl}
N.~Jaleeli, L.~VanSlyck, D.~Ewart, L.~Fink, and A.~Hoffmann, ``{Understanding
  automatic generation control},'' \emph{IEEE Transactions on Power Systems},
  vol.~7, no.~3, pp. 1106--1122, 1992.

\bibitem{2007NERCVer.:l.0.2}
``{NERC Reliability Concepts Ver.:l.0.2},'' Tech. Rep., 2007.

\bibitem{Kundur1994PowerControl}
P.~S. Kundur, N.~Balu, and M.~Lauby, \emph{{Power System Stability and
  Control}}, 1994.

\bibitem{Rebours2007AFeatures}
Y.~G. Rebours, D.~S. Kirschen, M.~Trotignon, and S.~Rossignol, ``{A Survey of
  Frequency and Voltage Control Ancillary Services—Part I: Technical
  Features},'' \emph{IEEE Transactions on Power Systems}, vol.~22, no.~1, pp.
  350--357, 2 2007.

\bibitem{Pirbazari2010AncillaryWorld}
A.~M. Pirbazari, ``{Ancillary services definitions, markets and practices in
  the world},'' in \emph{2010 IEEE/PES Transmission and Distribution Conference
  and Exposition: Latin America (T{\&}D-LA)}.\hskip 1em plus 0.5em minus
  0.4em\relax IEEE, 11 2010, pp. 32--36.

\bibitem{Grainger1985Volt/VarProblem}
J.~Grainger and S.~Civanlar, ``{Volt/Var Control on Distribution Systems with
  Lateral Branches Using Shunt Capacitors and Voltage Regulators Part I: The
  Overall Problem},'' \emph{IEEE Transactions on Power Apparatus and Systems},
  vol. PAS-104, no.~11, pp. 3278--3283, 11 1985.

\bibitem{Parikh2013FaultGOOSE}
P.~Parikh, I.~Voloh, and M.~Mahony, ``{Fault location, isolation, and service
  restoration (FLISR) technique using IEC 61850 GOOSE},'' in \emph{2013 IEEE
  Power {\&} Energy Society General Meeting}.\hskip 1em plus 0.5em minus
  0.4em\relax IEEE, 2013, pp. 1--6.

\bibitem{Habib2017Multi-Agent-BasedRestoration}
H.~F. Habib, T.~Youssef, M.~H. Cintuglu, and O.~A. Mohammed,
  ``{Multi-Agent-Based Technique for Fault Location, Isolation, and Service
  Restoration},'' \emph{IEEE Transactions on Industry Applications}, vol.~53,
  no.~3, pp. 1841--1851, 5 2017.

\bibitem{Castello2015ABus}
P.~Castello, P.~Ferrari, A.~Flammini, C.~Muscas, P.~A. Pegoraro, and
  S.~Rinaldi, ``{A Distributed PMU for Electrical Substations With Wireless
  Redundant Process Bus},'' \emph{IEEE Transactions on Instrumentation and
  Measurement}, vol.~64, no.~5, pp. 1149--1157, 5 2015.

\bibitem{Antonova2011CommunicationAutomation}
G.~Antonova, L.~Frisk, and J.-C. Tournier, ``{Communication redundancy for
  substation automation},'' in \emph{2011 64th Annual Conference for Protective
  Relay Engineers}.\hskip 1em plus 0.5em minus 0.4em\relax IEEE, 4 2011, pp.
  344--355.

\bibitem{Monticelli2000}
A.~Monticelli, ``{Electric power system state estimation},'' \emph{Proceedings
  of the IEEE}, vol.~88, no.~2, pp. 262--282, 2000.

\bibitem{Debs1970ASystem}
A.~Debs and R.~Larson, ``{A Dynamic Estimator for Tracking the State of a Power
  System},'' \emph{IEEE Transactions on Power Apparatus and Systems}, vol.
  PAS-89, no.~7, pp. 1670--1678, 9 1970.

\bibitem{Capitanescu2007ContingencyFlow}
F.~Capitanescu, M.~Glavic, D.~Ernst, and L.~Wehenkel, ``{Contingency Filtering
  Techniques for Preventive Security-Constrained Optimal Power Flow},''
  \emph{IEEE Transactions on Power Systems}, vol.~22, no.~4, pp. 1690--1697, 11
  2007.

\bibitem{Wibowo2014SecurityControl}
R.~S. Wibowo, T.~P. Fathurrodli, O.~Penangsang, and A.~Soeprijanto, ``{Security
  constrained optimal power flow incorporating preventive and corrective
  control},'' in \emph{2014 Electrical Power, Electronics, Communicatons,
  Control and Informatics Seminar (EECCIS)}.\hskip 1em plus 0.5em minus
  0.4em\relax IEEE, 8 2014, pp. 29--34.

\bibitem{FengDong2012PracticalFlow}
{Feng Dong}, ``{Practical applications of Preventive Security Constrained
  Optimal Power Flow},'' in \emph{2012 IEEE Power and Energy Society General
  Meeting}.\hskip 1em plus 0.5em minus 0.4em\relax IEEE, 7 2012, pp. 1--5.

\bibitem{Abido2006MultiobjectiveProblem}
M.~Abido, ``{Multiobjective evolutionary algorithms for electric power dispatch
  problem},'' \emph{IEEE Transactions on Evolutionary Computation}, vol.~10,
  no.~3, pp. 315--329, 6 2006.

\bibitem{Gaing2003ParticleConstraints}
Z.-L. Gaing, ``{Particle swarm optimization to solving the economic dispatch
  considering the generator constraints},'' \emph{IEEE Transactions on Power
  Systems}, vol.~18, no.~3, pp. 1187--1195, 8 2003.

\bibitem{Chowdhury1990ADispatch}
B.~Chowdhury and S.~Rahman, ``{A review of recent advances in economic
  dispatch},'' \emph{IEEE Transactions on Power Systems}, vol.~5, no.~4, pp.
  1248--1259, 1990.

\bibitem{FederalEnergyRegulatoryCommission2006SecurityRecommendations}
{Federal Energy Regulatory Commission}, ``{Security constrained economic
  dispatch: definition, practices, issues and recommendations},'' Tech. Rep.,
  2006.

\bibitem{Monticelli1987Security-ConstrainedRescheduling}
A.~Monticelli, M.~V.~F. Pereira, and S.~Granville, ``{Security-Constrained
  Optimal Power Flow with Post-Contingency Corrective Rescheduling},''
  \emph{IEEE Transactions on Power Systems}, vol.~2, no.~1, pp. 175--180, 1987.

\bibitem{Vargas1993ADispatch}
L.~Vargas, V.~Quintana, and A.~Vannelli, ``{A tutorial description of an
  interior point method and its applications to security-constrained economic
  dispatch},'' \emph{IEEE Transactions on Power Systems}, vol.~8, no.~3, pp.
  1315--1324, 1993.

\bibitem{Elacqua1982SecurityPool}
A.~Elacqua and S.~Corey, ``{Security Constrained Dispatch at the New York Power
  Pool},'' \emph{IEEE Transactions on Power Apparatus and Systems}, vol.
  PAS-101, no.~8, pp. 2876--2884, 8 1982.

\bibitem{Yuan2011ModelingSystems}
Y.~Yuan, Z.~Li, and K.~Ren, ``{Modeling Load Redistribution Attacks in Power
  Systems},'' \emph{IEEE Transactions on Smart Grid}, vol.~2, no.~2, pp.
  382--390, 6 2011.

\bibitem{Guo2001GlobalSystems}
Y.~Guo, D.~Hill, and Y.~Wang, ``{Global transient stability and voltage
  regulation for power systems},'' \emph{IEEE Transactions on Power Systems},
  vol.~16, no.~4, pp. 678--688, 2001.

\bibitem{Apostolopoulou2014AutomaticTime}
D.~Apostolopoulou, P.~W. Sauer, and A.~D. Dominguez-Garcia, ``{Automatic
  Generation Control and Its Implementation in Real Time},'' in \emph{2014 47th
  Hawaii International Conference on System Sciences}.\hskip 1em plus 0.5em
  minus 0.4em\relax IEEE, 1 2014, pp. 2444--2452.

\bibitem{1970IEEESystems}
``{IEEE Standard Definitions of Terms for Automatic Generation Control on
  Electric Power Systems},'' \emph{IEEE Transactions on Power Apparatus and
  Systems}, vol. PAS-89, no.~6, pp. 1356--1364, 7 1970.

\bibitem{Yoshida2000AAssessment}
H.~Yoshida, K.~Kawata, Y.~Fukuyama, S.~Takayama, and Y.~Nakanishi, ``{A
  particle swarm optimization for reactive power and voltage control
  considering voltage security assessment},'' \emph{IEEE Transactions on Power
  Systems}, vol.~15, no.~4, pp. 1232--1239, 2000.

\bibitem{Roytelman1995Volt/varSystem}
I.~Roytelman, B.~Wee, and R.~Lugtu, ``{Volt/var control algorithm for modern
  distribution management system},'' \emph{IEEE Transactions on Power Systems},
  vol.~10, no.~3, pp. 1454--1460, 1995.

\bibitem{1997ProposedFACTS}
``{Proposed terms and definitions for flexible AC transmission system
  (FACTS)},'' \emph{IEEE Transactions on Power Delivery}, vol.~12, no.~4, pp.
  1848--1853, 1997.

\bibitem{Phadke1993SynchronizedSystems}
A.~Phadke, ``{Synchronized phasor measurements in power systems},'' \emph{IEEE
  Computer Applications in Power}, vol.~6, no.~2, pp. 10--15, 4 1993.

\bibitem{DeLaRee2010SynchronizedSystems}
J.~De~La~Ree, V.~Centeno, J.~S. Thorp, and A.~G. Phadke, ``{Synchronized Phasor
  Measurement Applications in Power Systems},'' \emph{IEEE Transactions on
  Smart Grid}, vol.~1, no.~1, pp. 20--27, 6 2010.

\bibitem{Mackiewicz2006OverviewBenefits}
R.~Mackiewicz, ``{Overview of IEC 61850 and Benefits},'' in \emph{2006 IEEE PES
  Power Systems Conference and Exposition}.\hskip 1em plus 0.5em minus
  0.4em\relax IEEE, 2006, pp. 623--630.

\bibitem{Elgargouri2015IECSecurity}
A.~Elgargouri, R.~Virrankoski, and M.~Elmusrati, ``{IEC 61850 based smart grid
  security},'' in \emph{2015 IEEE International Conference on Industrial
  Technology (ICIT)}.\hskip 1em plus 0.5em minus 0.4em\relax IEEE, 3 2015, pp.
  2461--2465.

\bibitem{Xu2013TowardsEnvironment}
J.~Xu, C.-W. Yang, G.~Zhabelova, S.~Berber, and V.~Vyatkin, ``{Towards
  implementation of IEC 61850 GOOSE messaging in IEC 61499 environment},'' in
  \emph{2013 11th IEEE International Conference on Industrial Informatics
  (INDIN)}.\hskip 1em plus 0.5em minus 0.4em\relax IEEE, 7 2013, pp. 464--470.

\bibitem{Sidhu2011ConfigurationGOOSE}
T.~Sidhu, M.~Kanabar, and P.~Parikh, ``{Configuration and performance testing
  of IEC 61850 GOOSE},'' in \emph{2011 International Conference on Advanced
  Power System Automation and Protection}.\hskip 1em plus 0.5em minus
  0.4em\relax IEEE, 10 2011, pp. 1384--1389.

\bibitem{salmon2009mitigating}
D.~Salmon, M.~Zeller, A.~Guzm{\'a}n, V.~Mynam, and M.~Donolo, ``Mitigating the
  aurora vulnerability with existing technology,'' in \emph{36th Annual western
  protection relay conference}, 2009.

\bibitem{Selvam2012AdvancedApplications}
C.~Selvam, K.~Srinivas, G.~Ayyappan, and M.~Venkatachala~Sarma, ``{Advanced
  metering infrastructure for smart grid applications},'' in \emph{2012
  International Conference on Recent Trends in Information Technology}.\hskip
  1em plus 0.5em minus 0.4em\relax IEEE, 4 2012, pp. 145--150.

\bibitem{Arian2011AdvancedArchitecture}
M.~Arian, V.~Soleimani, B.~Abasgholi, H.~Modaghegh, and N.~S. Gilani,
  ``{Advanced Metering Infrastructure System Architecture},'' in \emph{2011
  Asia-Pacific Power and Energy Engineering Conference}.\hskip 1em plus 0.5em
  minus 0.4em\relax IEEE, 3 2011, pp. 1--6.

\bibitem{Ullah2013ACommunications}
M.~N. Ullah, A.~Mahmood, S.~Razzaq, M.~Ilahi, R.~D. Khan, and N.~Javaid, ``{A
  Survey of Different Residential Energy Consumption Controlling Techniques for
  Autonomous DSM in Future Smart Grid Communications},'' 6 2013.

\bibitem{Mohsenian-Rad2010AutonomousGrid}
A.-H. Mohsenian-Rad, V.~W.~S. Wong, J.~Jatskevich, R.~Schober, and
  A.~Leon-Garcia, ``{Autonomous Demand-Side Management Based on Game-Theoretic
  Energy Consumption Scheduling for the Future Smart Grid},'' \emph{IEEE
  Transactions on Smart Grid}, vol.~1, no.~3, pp. 320--331, 12 2010.

\bibitem{Mohsenian-Rad2010OptimalEnvironments}
A.-H. Mohsenian-Rad and A.~Leon-Garcia, ``{Optimal Residential Load Control
  With Price Prediction in Real-Time Electricity Pricing Environments},''
  \emph{IEEE Transactions on Smart Grid}, vol.~1, no.~2, pp. 120--133, 9 2010.

\bibitem{Cohen1987AnControl}
A.~I. Cohen, J.~W. Patmore, D.~H. Oglevee, R.~W. Berman, L.~H. Ayers, and J.~F.
  Howard, ``{An Integrated System for Residential Load Control},'' \emph{IEEE
  Transactions on Power Systems}, vol.~2, no.~3, pp. 645--651, 1987.

\bibitem{Lin2013OnGrid}
J.~Lin, W.~Yu, D.~Griffith, X.~Yang, G.~Xu, and C.~Lu, ``{On Distributed Energy
  Routing Protocols in the Smart Grid}.''\hskip 1em plus 0.5em minus
  0.4em\relax Springer, Heidelberg, 2013, pp. 143--159.

\bibitem{LasseterMicroGrids}
R.~Lasseter, ``{MicroGrids},'' in \emph{2002 IEEE Power Engineering Society
  Winter Meeting. Conference Proceedings (Cat. No.02CH37309)}, vol.~1.\hskip
  1em plus 0.5em minus 0.4em\relax IEEE, pp. 305--308.

\bibitem{Hatziargyriou2007Microgrids}
N.~Hatziargyriou, H.~Asano, R.~Iravani, and C.~Marnay, ``{Microgrids},''
  \emph{IEEE Power and Energy Magazine}, vol.~5, no.~4, pp. 78--94, 7 2007.

\bibitem{Jia2012ImpactsOperations}
L.~Jia, R.~J. Thomas, and L.~Tong, ``{Impacts of Malicious Data on Real-Time
  Price of Electricity Market Operations},'' in \emph{2012 45th Hawaii
  International Conference on System Sciences}.\hskip 1em plus 0.5em minus
  0.4em\relax IEEE, 1 2012, pp. 1907--1914.

\bibitem{sinha2017review}
A.~Sinha, P.~Malo, and K.~Deb, ``A review on bilevel optimization: From
  classical to evolutionary approaches and applications,'' \emph{IEEE
  Transactions on Evolutionary Computation}, vol.~22, no.~2, pp. 276--295,
  2017.

\bibitem{unCTC}
UN-CTC, ``United nations: The protection of critical infrastructure against
  terrorist attacks: Compendium of good practices compiled,'' 2018.

\bibitem{itu}
\BIBentryALTinterwordspacing
ITU. (2019) National cybersecurity strategies repository. [Online]. Available:
  \url{https://www.itu.int/en/about/Pages/default.aspx}
\BIBentrySTDinterwordspacing

\bibitem{protiviti}
\BIBentryALTinterwordspacing
Protiviti. (2017) China’s cybersecurity law and its impacts - key
  requirements businesses need to understand to ensure compliance. [Online].
  Available:
  \url{https://www.protiviti.com/CN-en/insights/china-cybersecurity-law-and-impacts}
\BIBentrySTDinterwordspacing

\bibitem{pillai2017indian}
R.~Pillai, R.~Sarngapani, and H.~Thukral, ``Indian manual for cyber security in
  power systems,'' \emph{CIRED-Open Access Proceedings Journal}, vol. 2017,
  no.~1, pp. 2787--2790.

\bibitem{bac}
J.~P. Diet. (2014) The basic act on cybersecurity, act no. 104.

\bibitem{lewis2015us}
J.~A. Lewis, \emph{US-Japan Cooperation in Cybersecurity}.\hskip 1em plus 0.5em
  minus 0.4em\relax Center for Strategic \& International Studies, 2015.

\bibitem{ter2018singapore}
C.~S.~A. of~Singapore, ``Singapore's cybersecurity strategy,'' 2016.

\bibitem{AEMO}
A.~E.~M. Operator, ``Summary report into the cyber security preparedness of the
  national and wa wholesale electricity markets,'' 2018.

\bibitem{africa}
\BIBentryALTinterwordspacing
Symantec. (2016) Cyber crime \& cyber security trends in africa. [Online].
  Available:
  \url{https://www.symantec.com/content/dam/symantec/docs/reports/cyber-security-trends-report-africa-interactive-en.pdf}
\BIBentrySTDinterwordspacing

\bibitem{ukparliament}
ITRE, ``Cyber security strategy for the energy sector: Directorate general for
  internal policies, policy department a: economic and scientific policy,
  european parliament,'' 2016.

\end{thebibliography}

\end{document}